\documentclass[%
 aip,
rsi,%
 amsmath,amssymb,
 reprint,%
]{revtex4-1}
 
\usepackage{graphicx}% Include figure files
\usepackage{dcolumn}% Align table columns on decimal point
\usepackage[utf8]{inputenc}
\usepackage{bm}% bold math
\usepackage{tabularx}
\usepackage{amsmath}
\usepackage{natbib}
\usepackage{siunitx}
\usepackage{color}

\begin{document}

%\preprint{AIP/123-QED}

\title{Impact of Network Topology on the Stability of DC Microgrids}% Force line breaks with \\
%\thanks{Footnote to title of article.}

\author{J. F. Wienand}
\email{j.wienand@physik.uni-muenchen.de}
\affiliation{ 
Faculty of Physics, Ludwig-Maximilians-Universität München, Schellingstraße 4, 80799 Munich, Germany.
}
 \author{D. Eidmann}
 \email{david.eidmann@stud.tu-darmstadt.de}
\affiliation{ 
Department of Civil and Environmental Engineering Sciences, Technische Universität Darmstadt,\\ Karolinenplatz 5, 64283 Darmstadt, Germany.
}

\author{J. Kremers}
\affiliation{
Laboratory of Geo-information Science and Remote Sensing, Wageningen University and Research, Droevendaalsesteeg 3, 6708PB Wageningen, The Netherlands.
}

\author{J. Heitzig}

\author{F. Hellmann}
\affiliation{Potsdam Institute for Climate Impact Research, PO Box 60 12 03, Potsdam 14412, Germany.}

\author{J. Kurths}
\affiliation{Potsdam Institute for Climate Impact Research, PO Box 60 12 03, Potsdam 14412, Germany.}
\affiliation{Department of Physics, Humboldt University of Berlin, Newtonstrasse 15, Berlin 12489, Germany.}

\date{\today}% It is always \today, today,

\begin{abstract}
%Due to the increase in semiconductor technologies that rely on DC by fundamental physics, DC power grids gain importance.

%
%Valid PACS numbers may be entered using the \verb+\pacs{#1}+ command.

We probe the stability of Watts--Strogatz DC microgrids, in which droop-controlled producers and constant power load consumers are homogeneously distributed and obey Kirchhoff's circuit laws. The concept of survivability is employed to evaluate the system's response to Dirac delta voltage perturbations at single nodes. A fixed point analysis of the power grid model yields that there is only one relevant attractor. Using a set of simulations with random networks we investigate correlations between survivability and three topological network measures: the share of producers in the network
%Jobst: is the following what you mean?
and the degree and the average neighbour degree of the perturbed node. 
Depending on the imposed voltage and current limits, the stability is optimized for low node degrees or a specific share of producers. Based on our findings, we provide an insight into the local dynamics of the perturbed system and derive explicit guidelines for the design of resilient DC power grids.
\end{abstract}

\keywords{Microgrids, Direct Current, Stability, Survivability, Topology}%Use showkeys class option if keyword
\maketitle

\begin{quotation}
% LEDs, transistors, solar cells - these important flagships of semiconductor technology are DC (Direct Current) devices in an world of AC (Alternating Current) power distribution. To avoid lossy conversion between both forms of electric energy, the idea of purely DC based power grids is regaining popularity. Since stability is the most important feature of a power grid, our study investigates the physical properties of such DC networks in terms of stability. Based on a numerical implementation of classical electrodynamics, we investigate what properties make a DC network stable and resilient against perturbations. Our study derives explicit construction rules for the power grid to avoid blackouts and failures due to operation limits being exceeded.

More than a century ago Nicola Tesla won a victory over Thomas Edison in the \emph{war of currents} and alternating current (AC) became the global standard for power grids \citep{mcpherson_war_2012}. Thanks to the availability of cheap AC transformers, electric energy can be transferred over thousands of kilometers through high voltage transmission lines. However, since the invention of the transistor, more and more devices are based on semiconductor technology \citep{brinkman_history_1997} and, as imposed by fundamental physics, only work with Direct Current (DC). These comprise not only every computer or LED but also environmentally friendly energy sources such as solar cells. At the same time the development of efficient power electronics which allow converting between DC voltage levels makes it possible to consider DC power grids again. Therefore, in the face of the digital transformation and this century's need for renewable energies, the idea of DC power grids is reviving and regaining significance for the future distribution of electric energy \citep{dragicevic_dc_2015}, in particular in the context of microgrids \citep{kakigano_low-voltage_2010}. We simulate the electrodynamics of such small-scale networks and probe how they react to certain perturbations. Our results identify topological motives which contribute to stability and will help to design robust DC microgrids.

\end{quotation} 

%-------------------------
%More than one century ago Nicola Tesla won victory over Thomas Edison in the "war of currents" and alternating current (AC) became the global standard for power grids. What leveraged this technique the most, was the availability of cheap AC transformers which allowed to transfer electric energy over thousands of kilometers at high voltages. However, since the invention of the transistor, more and more devices are based on semiconductor technology and, by fundamental physics, only work with DC. In particular, this comprises not only every computer and every LED but also environmentally friendly energy resources such as solar cells. Yet, as modern transmission lines mostly use AC, an increasing numbers of AC-DC-converters is needed, which cause significant power losses. Thus, in the face of the advancing digital revolution and the transformation towards renewable energies, the idea of pure DC power grids is reviving and regaining significance for the future distribution of electric energy. %remote communities 

\section{Introduction}
A power grid's essential task is to ensure a reliable and stable power supply. It must be guaranteed that local events at particular nodes in the network's infrastructure, such as a short circuit or turning on a device, do not entail a collapse of the whole system. Instead, the network should compensate the perturbation as fast and effectively as possible so that normal operation is restored. The transgression of critical current and voltage values during the system's response, as well as a collapse into a different undesirable state (e.g.\ a blackout) must be strictly avoided. This raises the question how a power grid should be designed to make it robust and capable of surviving various kinds of incidents.

Besides the properties of electronic devices and transmission lines, it is the network topology of the power grid that can strongly influence the grid's stability. Grid operators routinely perform detailed studies of their concrete systems, but by now there also exists a considerable body of work on the interplay of dynamical stability and network structure. One approach is to analytically analyse properties of the linearized AC power grid equations subject to a variety of perturbances, e.g. \cite{grunberg2016performance, 10.1371/journal.pone.0213550, poolla2017optimal, tyloo2018robustness, tamrakar2018propagation, guo2018graph, Zhangeaav1027,  plietzsch2019generalized, zhang2019dynamic}. Another very successful approach to understanding the non-linear equations, and in particular towards understanding which network motives and structures are stabilizing or destabilizing in a power grid context has proceeded through probabilistic stability notions like basin stability and survivability \cite{menck_how_2014, Schultz_2014, auer2016impact, hellmann2016survivability, nitzbon2017deciphering, auer2017stability, hellmann2018network, kim2016building, wolff2018power}. For instance, it has been found that the stability of AC power grids is undermined by dead ends in the network structure\cite{menck_how_2014}. Subsequent works refined this in many directions, for example by studying not just individual grids but an ensemble of realistic topologies \cite{Schultz2014growth}, which allowed meaningful statistics on a variety of structures. Using this ensemble, it was possible to identify specific topological structures that trigger the existence of accessible new limit cycles \cite{nitzbon2017deciphering}.

When it comes to understanding the impact of network topology on the full non-linear system, much of what has been investigated for AC in this probabilistic manner, is still unclear for DC power grids. Recent years have seen a number of works on control strategies and consensus algorithms for DC systems\cite{dragicevic_dc_2016, zhao2015distributed, meng2017review, de2018power, cucuzzella2018robust, trip2019distributed}, or more specifically HVDC systems \cite{andreasson2014distributed, zonetti2017tool}. Further a considerable amount of work exists for single bus DC microgrids \cite{dragicevic_dc_2016, dragicevic_dc_2015}. However, for no-communication, networked (multi-terminal) DC systems the resilience of the underlying distributed power dynamics remained fairly understudied, especially outside the linear regime \cite{anand2012reduced}. Hence, in this paper, we investigate the short time scale stability of DC power grids with non-linear loads in dependence on certain topological properties of the underlying network. The systems are inspired by the application in extremely low technology scenarios encountered in the swarm electrification in rural areas of developing nations\cite{swarm2015}.
%\textcolor{red}{In view of the large variety of possible grid sizes, we focus here on the important special case of rather small DC microgrids like the ones currently being deployed in some developing countries (ref?).}

\begin{itemize}

\item The model\cite{strenge_stability_2017} describes classical electrodynamics on a complex network of  uniformly distributed consumers and producers subject to communication free minimal control. It is studied analytically to show that normal operation is the only relevant attractor, which implies that stability measures made for multistable systems (like basin stability) are not helpful here.
\item To find the most stable network topology, simulations are carried out using a probabilistic scheme. 
%Jobst: colons are rarely used in English
Randomized Watts--Strogatz \cite{watts_collective_1998} networks are perturbed locally by instant Dirac-delta-like voltage jumps at a single randomly selected node. The system's response is then evaluated in terms of a stability measure which focuses on transient dynamics, \emph{survivability} \cite{hellmann_survivability_2016}.
\item The survivability is then correlated with three important topological network measures. We identify the share of producers in the power grid and the perturbed node's degree as primary indicators of a DC power grid's stability.
\item Finally, our findings are further enriched with physical intuition and traced back to short-range interactions between current and voltage dynamics.
\end{itemize}
Our model is stylized to be able to focus on system-wide features  and aims to complement electrical engineering studies that tend to focus on more detail. The structure and the parameters of the simulated networks \cite{strenge_stability_2017} were chosen to resemble microgrids as found at remote villages without power supply in a swarm electrification approach \cite{swarm2015}. In the first place such networks are built for low-power applications such as electric lighting. Although our results are obtained using this particular case, they are meant to demonstrate general motives inherent to the physics of DC power grids.

\section{The Model}
\subsection{Definition and Parameters}
%MENTION DAMPED HARMONIC OSCILLATOR = CONVERGING SPIRAL 
Our model of a unipolar DC power grid is defined by a set of differential equations which apply Kirchhoff's circuit laws to a network graph. Edges represent power lines while nodes constitute consumer or producer units. Power lines possess both resistance $R$ and inductance $L$ and connect adjacent nodes by carrying a current $i$. The network is represented by a directed graph to unambiguously define the current direction in each edge. Each node is equipped with a capacitance $C$ and operates at a voltage labelled $v_P$ for producers and $v_C$ for consumers, respectively. In addition, the capacitor is put in parallel with either a droop-controlled voltage generator in the case of producers (droop coefficient $K$ and targeted reference voltage $v_{ref}$) or an electrical load that dissipates energy at a constant power $P_{C} < 0$ (the negative sign follows the `generator convention') in the case of consumers. Consumers are modeled as constant power loads to account for the power-maintaining effect of power electronic converters (DC/DC and DC/AC) which are needed to meet the voltage requirements of a load \cite{singh_constant_2017}. The rationale for assuming Droop control for producers is to use a simple technique which
achieves the basic control goal of maintaining a reference voltage in the power grid, as it has been done previously \cite{anand2012reduced, strenge_stability_2017}. More sophisticated designs \cite{gavriluta2014decentralized} are a matter of active research and are beyond the scope of this paper. Further, recent work mostly concerns the study of systems with communication infrastructure  \cite{de2018power, cucuzzella2018robust} which we do not want to assume here. 

Following the model proposed by Strenge et al. \cite{strenge_stability_2017}, the temporal evolution of node voltages (index $n$) and edge currents (index $l$) is then defined by

\begin{align}
&L \frac{ di_{l} }{dt} = - R\,i_{l}+\sum_{adj.\;n} v_{n} \, (-1)^{d(l, n)},\;\;\label{eq:system1}\\   
&C \frac{dv_{P,n}}{dt} = K\,(v_{ref}-v_{P,n})-\sum_{inc.\;l}i_{l}\,(-1)^{d(l, n)},\;\;\label{eq:system2}\\   
&C \frac{dv_{C,n}}{dt} = \frac{P_{C}}{v_{C,n}}-\sum_{inc.\;l}i_{l}\,(-1)^{d(l, n)},\;\;    \label{eq:system3}
\end{align}
with $adj.\,n$ being the adjacent nodes and %$adj.\,l$ being the adjacent
$inc.\,l$ being the incident %Jobst: this is the correct term here, I replaced it throughout the text
edges (power lines). Here, all edges, producer nodes and consumer nodes are assumed to be identical within each category. The sums collect the voltage (current) contributions from all adjacent nodes (incident edges) and add them together consistently with the correct sign, according to the direction of the connected edges. The exponent $d(l,n)$ takes the value $1$ if edge $l$ points \emph{towards} node $n$ and is zero otherwise. Expressed in words, the differential equations state that, on the one hand, changes in edge currents are caused by voltage gradients and Joule heating ($R\,i$). On the other hand, the node voltages respond to the net current of incident edges as well as either to a constant power load current ($P_{C} / v_{c}$) or to a droop-controlled power generation current $i_{gen,n} = K(v_{ref,P}-v_{P,n})$. This depends on whether the node is a consumer or a producer, respectively. For a network with $N_{C}$ consumers, $N_{P}$ producers, and $N_L$ connecting power lines, the full system in Eqns.\ (\ref{eq:system1}--\ref{eq:system3}) comprises $N_{C}$ + $N_{P}$ node equations and $N_{L}$ edge equations. Their mutual coupling is introduced by the sums over the adjacent nodes and incident edges and is determined by the underlying network topology.

\begin{figure}[htp!]
\includegraphics[width=0.53\textwidth]{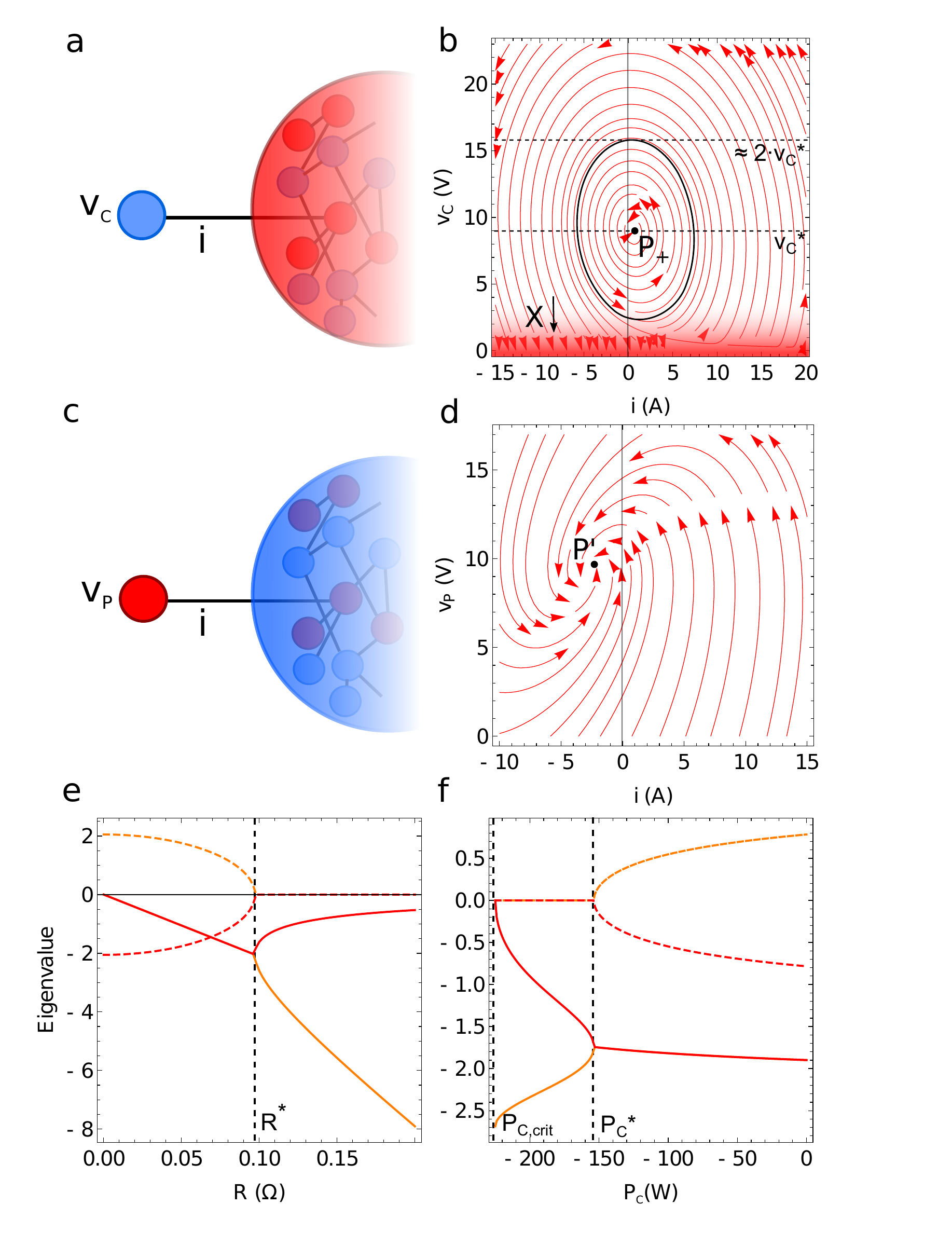}% Here is how to import EPS art
\caption{
(a--b) The DC power grid can be approximated with a two-node graph containing only the perturbed consumer node and a `supernode' representing the remainder of the network, summarized as one effective producer node with constant voltage $v_{P}$. The corresponding system of two differential equations (\ref{eq:reduced}) exhibits one exponentially stable fixed point $P_{+}$ (normal operation) and a voltage collapse $X$. The black curve indicates the border between the corresponding basins. (c--d) Analogously, for a producer node being perturbed, the remainder of the network acts as an effective consumer node with approximately constant voltage $v_{c}$. Then, (\ref{eq:reduced}) always converges to the only stable fixed point $P^{\prime}$ and normal operation is ensured. (e--f) Real (solid) and imaginary (dashed) parts of the eigenvalues (red and orange) of the Jacobi matrix at $P_{+}$, shown as a function of wire resistance $R$ and consumer power $P_{C}$. The attractor has a higher convergence speed for low $R$ and low $|P_C|$ and forms a stable
focus if $R<R^{\ast}$ or $P_{C}>P_{C}^{\ast}$.  $R^{\ast}$ and $P_{C}^{\ast}$ are the bifurcations which mark the transition from a stable node to a stable spiral (or vice versa).
}
\label{fig:eigenvalues} 
\end{figure}

\subsection{Equilibrium Voltage}

At normal operation, the total power fed into the system by $N_{P}$ producers and the power drawn by $N_{C}$ consumers must add up to zero. Neglecting Joule heating in the power lines we can write
\begin{equation}
N_{P} P_{P} + N_{C} P_{C} = 0. \label{eq:eqilibrium}
\end{equation}
Here, $P_{P} = \sum_{n} P_{P,n}/N_{P} $ denotes the average power production per producer, with $P_{P,n} = v_{P,n} \,i_{gen,n}$ being the product of the producer's node voltage $v_{P,n}$ and its power generating current $i_{gen,n} = K (v_{ref}-v_{P,n})$. If consumers and producers are uniformly distributed in all regions of the network, i.e.\  without forming clusters, all node voltages will equilibrate close to the overall average voltage $v_{eq}$, so that $v_{P,n} \approx v_{eq}$ and $P_{P} \approx v_{eq} K (v_{ref}-v_{eq}).$ Then (\ref{eq:eqilibrium}) can be solved for $v_{eq}$,
\begin{equation}
v_{eq}(s) = \frac{v_{ref}}{2} + \sqrt{\frac{v_{ref}^2}{4} - \frac{1-s}{s} \frac{P_{C}}{K}},
\label{eq:eqvoltage}
\end{equation}
where $s = N_{P}/(N_{P}+N_{C})$ is the share of producers in the network. The solution in (\ref{eq:eqvoltage}) approaches $v_{ref}$ in the limit of producer saturation ($s=1$) and collapses for $s < s_{crit} = P_{C}/(P_{C}-K v_{ref}^2/4)$. In the latter case, the total energy generated at producer nodes falls short of serving the total energy demanded by consumers. Note that this attractive singularity is always present. For a sufficiently small voltage at a consumer node the $P_{C}/v_{C}$ term in \eqref{eq:system3} dominates and the system hits the singularity $v = 0$ in finite time. We will return to the possibility of this voltage collapse later.

\subsection{Two-Node Approximations}

The term $P_{C} / v_{C,n}$ in (\ref{eq:system3}) expresses the rigid attempt of a consumer node to draw energy at constant power, such that a lower node voltage $v_{C}$ is immediately compensated by drawing more current from incident  edges. With $v_{C}$ in the denominator, Eqns.\ (\ref{eq:system1}--\ref{eq:system3}) represent a non-linear system and thus, in combination with a graph-based power grid structure, drive the dynamics of the complex network. While a detailed theoretical study of the full system's dynamics is beyond the scope of this paper, grouping nodes into super-nodes with effective properties does allow to derive approximations for the overall behavior of the power grid. 

When the normal operation of the network is locally perturbed by a sudden voltage jump at one single node, the interaction of the affected node with the remainder of the network can threaten the stability of the power grid. Both the exceedance of particular current or voltage values and the relaxation into a different equilibrium (e.g.\ collapse), rank among the undesirable scenarios which may occur during the system's journey through the $2 (N_{P}+N_{c}+N_{L})$-dimensional phase space.
We assume that for sufficiently small perturbations or sufficiently large networks the average voltage in the remainder of the network will not change significantly in response to the local perturbation. From a physical point of view, the validity of this assumption arises from the node capacitors, which are put in parallel by the network structure, such that the total capacitance becomes very large. Thus, it is comparably easy for a local perturbation to change one node voltage, but very difficult to change node voltages all over the network far away from the perturbed node. Hence, we model the interaction of the perturbation with the power grid simply with a two-node network, in which the first node is the perturbed node and the second node summarizes the remainder of the network at fixed voltage. This infinite grid approximation is also typical in AC power grid studies \citep{kozhaya_multigrid-like_2002}. Although its range of validity will require further study, it makes the model analytically tractable and gives valuable insights consistent with our simulations.

\subsubsection{Perturbation at a Consumer Node}
\label{sec:the}
Applying this approximation to a perturbation at a consumer node (Fig.~\ref{fig:eigenvalues}a), the full system of Eqns.\ (\ref{eq:system1}--\ref{eq:system3}) reduces to only two differential equations,
\begin{eqnarray}
L \frac{ di }{dt} = v_{P}-v_{C}  - R i,\;\;\;\;
C \frac{dv_{C}}{dt} = i+\frac{P_{C}}{v_{C}}.\;\;
\label{eq:reduced}
\end{eqnarray}
(\ref{eq:reduced}) yields two fixed points $P_{\pm}=(i_{\pm}^{\ast},v_{C \pm}^{\ast})$ located at
\begin{eqnarray}
P_{\pm}=\left(\frac{v_{P} \mp \sqrt{4 P_{C} R + v_{P}^{2}}}{-2 R}, \frac{v_{P} \pm \sqrt{4 P_{C} R + v_{P}^{2}}}{2}\right).
\label{eq:fixedpoints}
\end{eqnarray}
As illustrated in Fig.~\ref{fig:eigenvalues}b for $R = \SI{0.09}{\ohm}$, $P_{C}=\SI{-5}{\watt}$, $C = \SI{1}{\farad}, L =\SI{1}{\henry}$ and $v_{P} = \SI{9}{\volt}$, the fixed point $P_{+}$ is exponentially stable. This is the desired equilibrium state of the system corresponding to normal operation. $P_{-}$, however, is located at a negative voltage and, thus, considered a non-physical solution.

In addition to these two fixed points, just like the full system, the reduced system (\ref{eq:reduced}) might collapse into $v_{C}=0$ which corresponds to a breakdown of the power grid ($X$ in Fig. \ref{fig:eigenvalues}b). As noted above, for $v_{C}\ll1$, the term $P_{C}/v_{C}$ in (\ref{eq:reduced}) with $P_{C}<0$ dominates causing a rapid voltage decrease from which the system does not recover. The controller tries to draw more power by lowering the voltage but can not generate sufficient current to reach the desired constant power load and hits the singularity at $v_C=0$ in finite time. The red region in Fig.\ \ref{fig:eigenvalues}b indicates the regime in which the voltage collapse occurs. 

% Although this collapse is inherent to \ref{eq:reduced}, it is not accessible with our model, as for voltages close to zero ($v\rightarrow0)$ a constant power load would have to draw infinite current due to $P_{C} = v i $.

The trajectory towards the attractor $P_{+}$ pictures a converging spiral (Fig.~\ref{fig:eigenvalues}b), hence it is a stable focus. This behavior is consistent with the pair of complex conjugated eigenvalues of the corresponding Jacobi matrix, as seen in Fig.~\ref{fig:eigenvalues}e~and~f for $R<R^{\ast}$ and $P_{C}>P_{C}^{\ast}$. At the bifurcations $R=R^{\ast}$ and $P_C=P_{C}^{\ast}$ the two eigenvalues become real and the focus of $P_{+}$ transforms into a stable node. The convergence rates of both spiral and node are given by the real part of the eigenvalues. Subsequently, the system is expected to approach $P_{+}$ faster for a lower wire resistance and less energy consumption at the consumer node. Otherwise, if the consumer power exceeds a critical value ($P_{C} < P_{C,crit}$), the square roots in (\ref{eq:fixedpoints}) are complex, the fixed points $P_{\pm}$ vanish and the voltage collapse remains the only attractor the system can converge to. %%correct?

Whether the power grid enters normal operation at $P_{+}$ or collapses ($X$) depends on the initial conditions $(i_{0}, v_{C,0})$ imposed by the type and the strength of the perturbation. For voltage perturbations, the particular form and position of the focus in state space (Fig.~\ref{fig:eigenvalues}b) allows to estimate the upper and lower limits of $v_{C}$, within which a perturbation does not lead to a collapse of the power grid: When a node is abruptly charged or discharged, the difference in energy $\Delta E_{C}=1/2\,C (v_{C,0}^{2}- {v_{C}^\ast}^{2})$ triggers a current $i$. However, after the injected energy has been transferred to the incident  power lines ($1/2\,L i^{2} = \Delta E_{C}$), the current does not stop immediately, but, while declining, continues to further discharge the consumer node almost by another energy quantity $\Delta E_{C}$, due to the inductance of the wire. This causes the first kick in the temporal evolution of $v_{c}$ to be delimited by approximately $[v_{C}^\ast-|v_{c,0} - v_{C}^\ast|,v_{C}^\ast+|v_{c,0} - v_{C}^\ast|]$, provided that the convergence rate to the attractor is not too high. Since the basin of attraction of the focus nearly touches zero consumer voltage, the perturbation must either bring the node voltage very close to zero or exceed twice the equilibrium voltage (\ref{eq:eqvoltage}) to leave the basin of $P_{+}$ and to entail the collapse X into the red region in Fig.~\ref{fig:eigenvalues}b. This is why stability measures based solely on basin size (basin stability) are not a helpful tool here and one has to consider the transient behaviour to distinguish different degrees of stability against non-infinitesimal perturbations.

\subsubsection{Perturbation at a Producer Node}

Analogously to the case when the perturbation strikes a consumer node, the simplified system for a producer node (Fig.~\ref{fig:eigenvalues}c) reads
%(vp - vc) - R i, -i + K (vref - vp)
\begin{align}
\label{eq:reduced2a}
&L \frac{ di }{dt} = v_{P}-v_{C}  - R i,\;\; \\
&C \frac{dv_{P}}{dt} = -i + K (v_{ref} - v_{P}).\;\;
\label{eq:reduced2b}
\end{align}
Eqns.\ (\ref{eq:reduced2a}--\ref{eq:reduced2b}) always converge to the only and exponentially stable fixed point 
\begin{equation}
P^{\prime} = \left(\frac{K (-v_{C} + v_{ref})}{1 + K R}, \frac{v_{C} + K R v_{ref}}{1 + K R}\right)
\label{eq:fixedpoint2},
\end{equation}
which, just like $P_{+}$, corresponds to the state of normal operation. Hence, in contrast to a perturbation at a consumer node, the power grid does not enter a different attractor if a producer node is perturbed (Fig.~\ref{fig:eigenvalues}d).%, unless in the full system (\ref{eq:system1}~-~\ref{eq:system3}) sufficiently large voltage amplitudes cause neighboring consumer nodes to collapse.
\\ \\
In summary, the study of equilibrium states and non-equilibrium dynamics suggests that for voltage perturbations applied to individual nodes, the power grid reliably returns to normal operation without collapsing into a different attractor. The collapse regime $X$ is only relevant under extreme conditions that lie beyond the interest of our investigation. This conclusion is valid for the reduced systems of equations (\ref{eq:reduced}) and (\ref{eq:reduced2a}--\ref{eq:reduced2b}) which approximate the full network. In the realm of AC power grids, there is a profusion of attractors, including anomalous ones that can not be understood in the reduced way \cite{nitzbon2017deciphering}. However, it appears that the same is not the case for the DC system since during all numerical tests involving a wide range of parameters no other stable attractors than those denoted above with $P_{+}$ or $P^{\prime}$ (normal operation) and the collapse $X$ have ever been observed in our model. Therefore, our assessment of DC power grids in terms of stability must be based on measures which evaluate the system's trajectory inside the only relevant basin of attraction of normal operation around $P_{+}$ or $P^{\prime}$, respectively. For this purpose, we apply the probabilistic concept of survivability\cite{hellmann_survivability_2016}, which is introduced in the following section.

\par
\section{Simulation Methods}

\begin{table}[htp!]
\begin{tabularx}{0.485 \textwidth}{ll}
\textbf{Network Parameters}  &\\\hline\\[-1.0em]
Network Model & Watts--Strogatz \\
Rewiring Probability & $[0, 1]$ \\
Number of Nodes ($N_C+N_P$)& $[10, 100]$\\ 
Mean Degree & $2,4,6$ \\\\
\textbf{Simulation Parameters}  &\\\hline\\[-1.0em]
Producer Share ($s$)& $[0, 1]$\\
Droop Coefficient ($K$)& $\SI{1}{\ohm^{-1}}$\\
Power Line Inductance ($L$)&  $0.237\cdot10^{-4}$\ \si{\henry}\\
%0.237 10\tothe{-4}
Node Capacitance ($C$)& \SI{0.01}{\farad} \\
Resistance ($R$)& \SI{0.0532}{\ohm}\\
Reference Voltage ($v_{ref})$& \SI{48}{\volt}\\
Consumer Load Power ($P_{C}$)& \SI{-12}{\watt}\\
Voltage Perturbation& [\SI{44}{\volt}, \SI{48}{\volt}] \\\\
\textbf{Survivability Limits}  &\\\hline\\[-1.0em]
Voltage & [\SI{44}{\volt}, \SI{48}{\volt}] \\
Current & [\SI{-9}{\ampere},  \SI{9}{\ampere}] \\
\end{tabularx}
\caption{Physical properties of the perturbed DC power grids including voltage and current limits used for the survivability measure.}
\label{tab:settings}
\end{table}

We simulated DC Power Grids on Watts--Strogatz networks by integrating Eqns.\ (\ref{eq:system1}--\ref{eq:system3}) with a step size of $\SI{20}{\micro\second}$. % (Each node represents a producer respective consumer and each edge a power line.)
For each simulation, the parameters of the Watts--Strogatz model (rewiring probability, number of nodes, mean degree) as well as the producer share were randomly chosen from fixed intervals (Tab.\  \ref{tab:settings}). Our parameter choice is an example for one particular DC microgrid. However, the same qualitative behaviour of the system was also observed for other values, including a variation of the node capacitance and the wire resistance over two orders of magnitude. The minimum number of nodes in a network was set to $10$ and the maximum number of nodes was $100$.
The mean degree did not exceed six to ensure a non-trivial and, thus, more realistic, not excessively connected network topology. Other network parameters not in the focus of this study (consumer power consumption, reference voltage, power line resistance, capacity and droop coefficient) were set to constant values \cite{strenge_stability_2017} (cf. Tab.\ \ref{tab:settings}).

Every simulation was initiated with zero current on all edges and with the reference voltage $v_{ref}=\SI{48}{\volt}$ at all nodes. To calculate the equilibrium state of normal operation, each grid was simulated for $\SI{0.2}{\second}$ without perturbation. Subsequently, one perturbation was applied to a single node, randomly chosen from the nodes of the network. The perturbation constitutes an instant voltage jump with a new voltage value randomly selected from the interval [\SI{44}{\volt}, \SI{48}{\volt}]. From this point on, the response of the system, e.g.\ all voltages and currents in the network, was recorded over time until it had converged back to its previous equilibrium state. This usually took up to $\SI{0.1}{\second}$. If currents or voltages exceed particular boundaries during the simulation, the simulation run is counted as \emph{not survived}, otherwise as \emph{survived}. The applied permissible voltage and current intervals read [\SI{44}{\volt}, \SI{48}{\volt}] and [\SI{-9}{\ampere}, \SI{9}{\ampere}], respectively, and were chosen to model a DC microgrid, which currently is the most common application of DC power grids \cite{dragicevic_dc_2016,backhaus_dc_2015,dragicevic_dc_2015}. Any state of the system inside the boundaries does, by far, not trigger a collapse scenario. The sequence of locally perturbing a random node and recording the response was repeated 100,000 times, creating a set of random simulations. 

The share of survived simulations is called the \emph{survivability} of this specific set. Survivability can be interpreted as the probability for a system to survive a random perturbation which does not kill the system instantaneously \cite{hellmann_survivability_2016}. Here, it is chosen as a primary measure, as it combines many physical effects into one quantification of stability. This allows the identification of the functionality-controlling parameters, prior to understanding all complex physical interactions of the model. %In this scope, survivability is used as the main measurement of network stability.
Desirable network parameters are those which lead to a high survivability of a set of networks. %Survivability was first introduced by Hellmann et al. \cite{hellmann} to define a measure that mets a system's nature to die when particular limits are exceeded. Especially for power grids that are bound to certain limitations, for example by fuses or bypass voltages, survivability is a simple measure for combining the complex interaction of different phenomena.

  \begin{figure*}[htp!]
\includegraphics[width=1\textwidth]{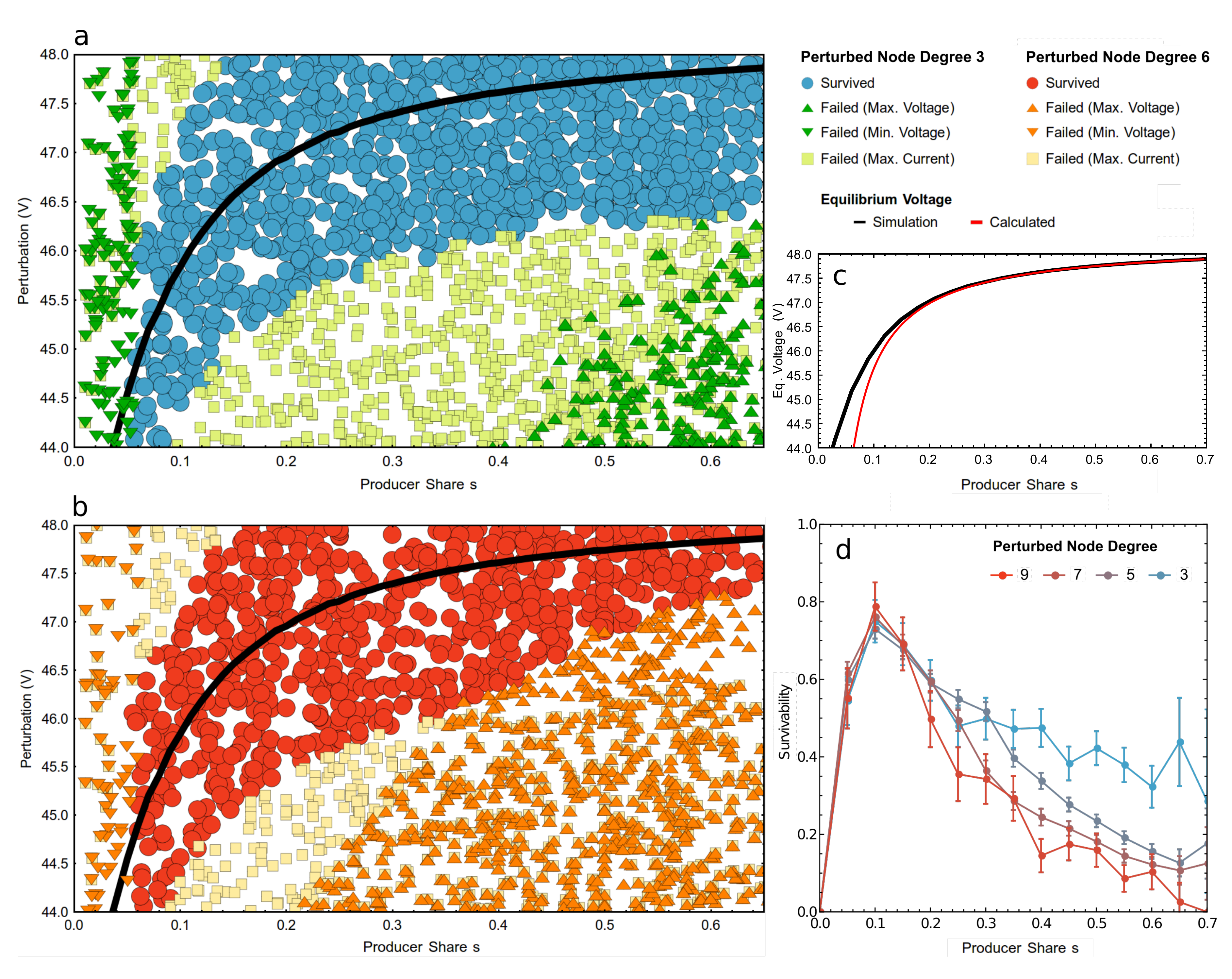}% Here is how to import EPS art
\caption{\label{fig:wide} %\textbf{Simulation results for perturbations at different node degrees.}
Randomly generated DC power grids on Watts--Strogatz graphs are locally perturbed and assessed in terms of survivability. (a) Each marker corresponds to the outcome of one simulation (survived or failed) with perturbations at nodes with degree 3. Plotted as a function of the perturbation voltage (ranging from $\SI{44}{\volt}$ to $\SI{48}{\volt}$) and the share of producers in the perturbed network, they form a \emph{band of survived simulations}: It surrounds the average equilibrium voltage (black line) and is mostly delimited by violations of the current boundary ($\SI{9}{\ampere}$). (b) For perturbed nodes with a higher degree of 6, the band is narrowed by violations of the upper voltage boundary $\SI{48}{\volt}$. (c) The analytically derived equilibrium voltage (\ref{eq:eqvoltage}) agrees well with the simulation. (d) Since the vertical height of the band of survived simulations corresponds to the survivability as a function of producer share, the chosen voltage boundaries cause the survivability to become maximal at a producer share of $12\pm3~\%$. Beyond this threshold, higher node degrees imply less stability.}% due to voltage failures featured in Fig. \ref{fig:simulation}}.
\end{figure*}

\section{Simulation Results}

\subsection{Average Equilibrium Voltage}
The average equilibrium voltage in the power grid is predicted to be related to the producer share through (\ref{eq:eqvoltage}). The simulated values match the theoretical prediction almost perfectly (Fig.\ 
\ref{fig:wide}c). Merely for low-producer shares the agreement deteriorates. For $s$ approaching $s_{crit}$, fewer producers feed more consumers and, due to an increased difference between consumer and producer voltages, the approximation $v_{eq}\approx v_{P}$ in the derivation of (\ref{eq:eqvoltage}) is not well fulfilled anymore. %really? 1

\subsection{Band of Survived Simulations}
Fig.~\ref{fig:wide}a and b depict the outcomes (survived or failed) of about 4000 simulations as a function of both the share of producers in the network and the induced voltage perturbation. With one marker per simulation, the data is sorted by the degree of the perturbed node (3 in fig.~\ref{fig:wide}a  and 6 in fig.~\ref{fig:wide}b). As suggested in Section \ref{sec:the}, the average equilibrium voltage is of substantial importance for the perturbation interval within which simulations survive. From Fig.~\ref{fig:wide}a and b it is apparent that a crucial condition for survivability lies in a small perturbation magnitude relative to the average equilibrium voltage. Markers representing survived simulations form a band which is centered symmetrically around the simulated average equilibrium voltage line (black curve).% The vertical width of that band decreases for small producer shares. From an electrodynamic point of view, this narrowing can be attributed to the voltage consumption term in the consumer voltage equation (\ref{eq:system3}) of the power grid model. Under equilibrium conditions, $dv_{c}/dt=0$ and $\sum_{k} i_{k}=-\frac{P}{v_{c}}$ (\ref{eq:system3}). When $v_{c}$ is altered by an infinitesimally small perturbation $v_{c} \rightarrow v_{c} + \delta v_{c}$, the change in the sum of connected currents in first order approximation corresponds to
%\begin{equation}
%    \delta(\sum_{k} i_{k}) = P \; \frac{\delta  v_{c}}{v_{c}^{2}}
%    \label{eq:prop}
%\end{equation}
%according to variational calculus. 
%The more producers feed the network, the closer the node equilibrium voltages approach the globally fixed reference voltage of $48V$. As the current variation scales with $1/v_{c}^{2}$, a higher producer share is expected to gradually suppress the perturbation term and to widen up the power grid's tolerance towards stronger perturbations. In general, the width of the band of survivability is set by the upper current boundary ($10A$ in our simulations).

\subsection{Voltage and Current Failures}

The simulations which do not survive are those whose perturbation voltage lies too far away from the average equilibrium voltage (black curve), i.e.~beyond the edge of the band of survived simulations. Either the current or the voltage limits are violated during the system's temporal evolution. As apparent from Fig.~\ref{fig:wide}a and b, in the vast majority of simulations it is the current boundary of $\SI{9}{\ampere}$ which causes the power grid to fail and which confines the band of survived simulations to its narrow range around the average equilibrium voltage. Failures due to current violations occur independently of the perturbed node's degree.
In the regimes of high and very low producer shares, a violation of voltage limits is observed. Minimum voltage violations below a producer share of $4\,\%$ are attributed to the observation that the average equilibrium voltage falls below the lower voltage boundary. Maximum voltage violations are suppressed for small node degrees: Perturbed nodes with a degree of 6 (Fig.~\ref{fig:wide}b) are much more likely to exceed the upper voltage boundary compared to those with a lower node degree of 3 (Fig.~\ref{fig:wide}a). In the former case (Fig.~\ref{fig:wide}b), the sensitivity of the power grid towards voltage failures, even for perturbations very close to the equilibrium voltage, is so high that the band of survived simulations is narrowed on its bottom side at high producer shares.

\subsection{Survivability}

Applying the survivability measure to Fig.~\ref{fig:wide}a or b means counting the share of survived simulations (i.e.\  those simulations whose voltage and current values remained within their respective bounds) along the vertical axis for all producer shares on the horizontal axis. The result of this procedure is shown in Fig.~\ref{fig:wide}d which summarizes the relationship between a power grid's survivability and the perturbed node's degree as a function of producer share. %More pronounced than its dependency on node degree,T
The survivability is maximized for a particular producer share of about $12\,\%$. This maximum is a consequence of both the upper and the lower voltage boundaries which have been set to $\SI{48}{\volt}$ and $\SI{44}{\volt}$, respectively. As seen in Fig.~\ref{fig:wide}a and b, these limits crop the vertical width of the band at high and low producer shares. In these regimes, the share of survived simulations within the perturbation voltage interval under consideration is diminished and results in a lower survivability. 
Both trends combined entail a survivability maximum at an intermediate producer share value of about $12\%$. 
However, if the upper voltage limit is raised (lowered), so that a different proportion of the survivability band is cut off, the region of largest vertical width of the band, and thus the region of optimal survivability, broadens (narrows) asymmetrically to the right (left), i.e.\  to higher (lower) producer shares (not shown). % If, additionally, a higher minimum voltage limit is employed which cuts the survivability band at the bottom, the region of highest survivability inside the network parameter space receives further broadening shifts (not shown).
%For voltage limits far away from the margins of the band, the survivability is expected to become independent of the producer share.
Hence, the producer share parameter demonstrates how optimal network parameters (highest stability) depend on the boundaries of the survivability measure. This is why, in order to build the optimal DC power grid, one needs to know precisely the range of permissible voltage and current values. 
 
\begin{figure}[htp!]
\includegraphics[width=0.47\textwidth]{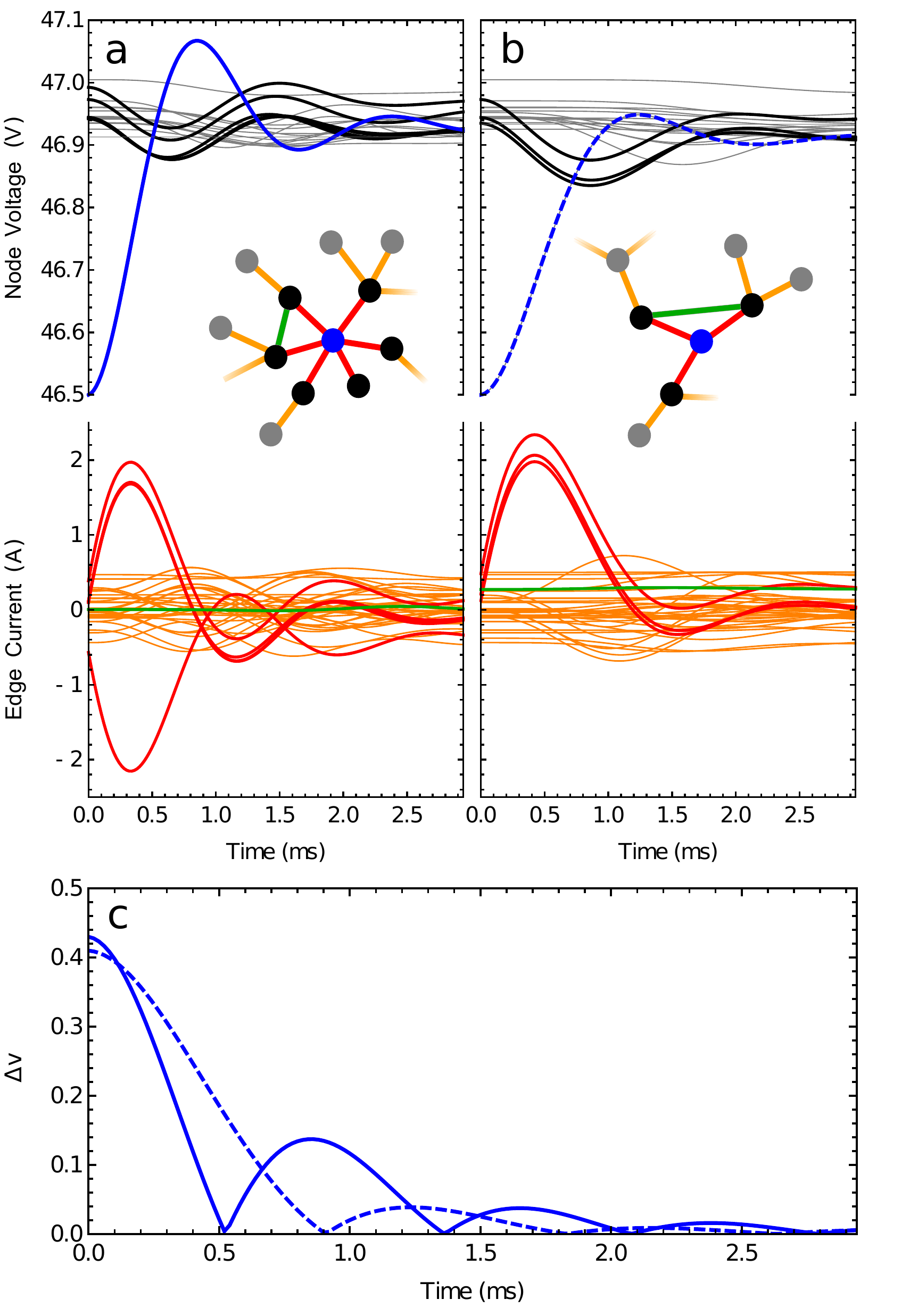}% Here is how to import EPS art
\caption{\label{fig:simulation} 
(a-b) More neighbours (black) cause the perturbed node (blue) to overcompensate the perturbation and to attain more extremal voltage values. This is apparent from comparing the temporal evolution of node voltages and edge currents in response to a perturbation (\SI{46.5}{\volt}) at a node with a high degree of 6 (a) and a low degree of 3 (b). Colors encode the proximity to the epicentre, as indicated in the schematic network graph (inset): The environment of the perturbed node (blue) is categorized into first-order edges (red), higher order edges (orange), first-order bridging edges (magenta). first-order neighbours (black) and higher-order neighbours (gray). The perturbation at $t=0$ causes the DC power grid to perform a damped wave, whose period length and damping coefficient is enhanced for the low-degree node. As the spreading of the electrical potential to adjacent neighbours is roughly isotropic, bridging edges remain unaffected by the perturbation. (c) Absolute deviation $\Delta v = |v - v_{eq}|$ of the perturbed node voltages from their equilibrium values (dashed: degree~3, normal: degree~6). 
}
\end{figure}

\subsection{Node Degree}
In order to explain the observed differences between perturbations at nodes with a lower and higher degree, we take a closer look at the perturbed node. In Fig.~\ref{fig:simulation} we juxtapose a voltage perturbation of $\SI{46.5}{\volt}$ at a consumer node with degree 3 (Fig.~\ref{fig:simulation}b) and at a consumer node with degree 6 (Fig.~\ref{fig:simulation}a). The insets picture the environments of the perturbed nodes (blue), including their incident  edges (red) and adjacent nodes (black) as well as second order neighbouring edges (yellow) and nodes (grey) and first-order bridging edges (green). The plotted curves in corresponding colors depict the voltage and current evolution over time after the out-of-equilibrium voltage perturbation at $t=0$. The sudden voltage jump triggers a damped oscillation of the node voltages and edge currents. 
%and a subsequent attempt of each node to reestablish the stable equilibrium voltage. 
While the voltage of the perturbed node (blue) starts oscillating immediately after the perturbation, higher order neighbours and power lines are affected with a delay, depending on the distance to the epicentre. In addition, also the amplitude of the oscillation is diminished further away from the perturbed node: These nodes are only affected indirectly and the intermediate nodes screen the perturbation. First-order bridging edges (green) do not exhibit any current oscillations, as adjacent nodes are equally distant from the perturbed node and, during the radially expanding perturbation wave, experience roughly the same electrical potential.

Contrasting the responses to perturbations at nodes with different degrees, one can see a pronounced attenuation of the voltage and current oscillations at the node with degree 3 (Fig.~\ref{fig:simulation}b). In particular, the perturbed node with higher degree experiences a faster voltage oscillation with an amplitude declining more slowly. This is due to the initial voltage drop being overcompensated by more current from more incident  power lines, which results in a subsequent voltage overshoot higher above the equilibrium voltage. Fig.~\ref{fig:simulation}c depicts the absolute deviation of the perturbed node's voltage from its equilibrium value $\Delta v = |v - v_{eq}|$ and particularly highlights this observation. Hence, some power grids do not survive the perturbation because the voltage overshoot might lead to a violation of voltage boundaries. This scenario is more probable when the average equilibrium voltage lies closer to the upper voltage boundary of the survivability measure, what is the case for a large share of producers in the network (cf.\ Fig.~\ref{fig:wide}c). Then, the survivability is diminished depending on the degree of the perturbed node (cf.\ Fig.~\ref{fig:wide}d).
%On average, the perturbed network simulations exceeding a maximum voltage of $48~V$ are those perturbed at higher degree nodes compared to the ones which do not exceed this voltage survivability threshold.
% On another note, the frequency of the voltage and current variations is higher for these nodes. This behaviour can be explained by the number of neighbouring nodes and edges. Since the node with a higher degree has more adjacing nodes and edges from which to draw energy, voltage and current levels change faster.

\begin{figure}[htp!]
\includegraphics[width=0.45\textwidth]{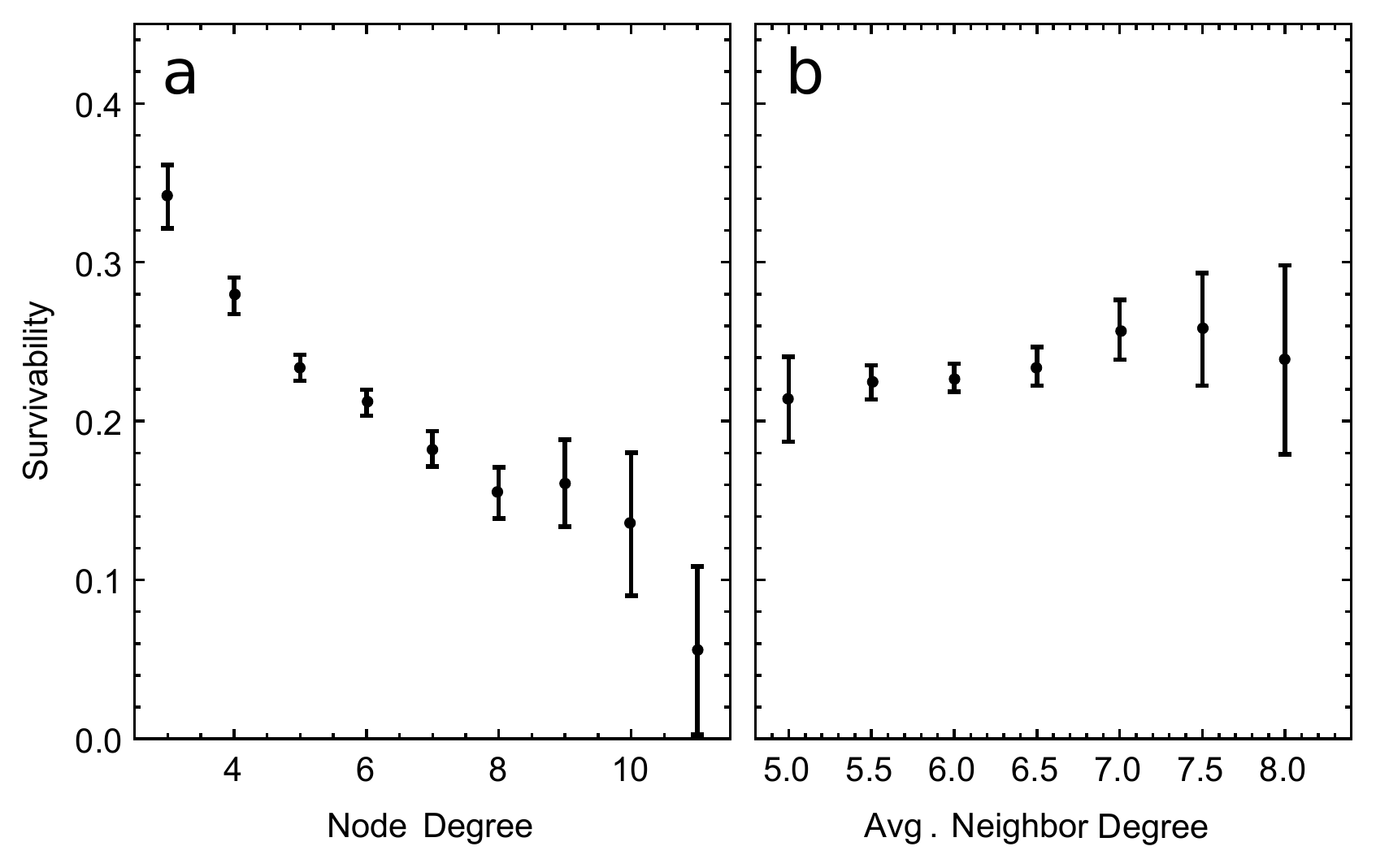}% Here is how to import EPS art
\caption{\label{fig:survivability}
As the perturbation response declines fast in both time and space, the fate of the power grid is decided within the radius of first-order neighbours of the perturbed node and during the first period length of voltage and current oscillations. Thus, for a producer share $s=0.5$, one observes a distinct correlation between the survivability and the perturbed node's degree (a), while higher-order measures, such as the perturbed node's average neighbour degree (b), do not exhibit this feature.
}
\end{figure}

The vulnerability of high degree nodes towards undesirable voltage peaks can also be understood from the eigenvalue profile depicted in Fig.~\ref{fig:eigenvalues}e. The more neighbours a node has, the more edges connect the perturbed node with the remainder of the network, summarized as one supernode (cf. Fig.~\ref{fig:eigenvalues}a and c). These edges are in parallel configuration, which, according to Kirchhoff's circuit laws, lowers the total resistance of the connection. From Fig.~\ref{fig:eigenvalues}e it is apparent that a lower resistance entails a lower convergence rate of the focus around the state of normal operation. As a consequence, the likelihood of striking more extremal voltage or current values in phase space is enhanced and the survivability decreases. 

Therefore, apart from the magnitude of the perturbation and the producer share (global network parameter) the degree of the perturbed node (local parameter) is another potent predictor for a DC power grid's survivability (Fig.~\ref{fig:survivability}a). Higher-order measures like the perturbed node's average neighbour degree do not exhibit such a correlation (Fig.~\ref{fig:survivability}b). In accordance with the black, grey and orange curves in Fig.~\ref{fig:simulation}a and b, this is because the perturbation is damped fast not only in time but also in space, mainly within the radius of first-order neighbours.

Interestingly, our results stand in contrast with stability studies regarding AC networks. Menck et al.\cite{menck_how_2014} investigated the stability of AC power grids using the global average of the single-node basin stability. They found a strong influence of the perturbed node's average neighbour degree on the power grid's stability, whereas they found close to no influence of the first-order node degree on the power grid's stability. However, in a different study\cite{nitzbon_deciphering_2017}, it was shown with survivability that at least for specific classes of nodes, the so called sprouts, the neighbour degree has also a significant impact on the dynamics, even leading to novel bifurcations.

%% DO WE NEED THE FOLLOWING PARAGRAPH
%Increased voltage oscillations for perturbations at nodes with higher degrees lead to a lower survivability for perturbations at such nodes (Fig.~\ref{fig:survivability}a).

\section{Concluding Remarks}
We have examined the stability of DC power grids responding to a single-node voltage perturbation. We believe such an incident to be representative for various other common power grid perturbations like a sudden load increase or a short circuit. Both trigger a voltage and current wave expanding from the epicentre into the remainder of the network.

The stability of a power grid is defined as its ability to return to normal operation after the perturbation. Trespassing certain thresholds jeopardizes this ability. For our DC microgrids, we have assumed that single-node voltages should neither fall below $\SI{44}{\volt}$ nor exceed 48V, nor should single-edge currents exceed $\SI{9}{\ampere}$. If a simulated perturbed network stays within the desirable regime it is counted as survived, otherwise as not survived. Thus, applied to a single case, the survivability expresses whether a power grid (network model) of a certain configuration survives a given perturbation. Additionally, it serves as a frequentist measure of the proportion of simulated network models with certain configuration parameters surviving a given perturbation.

The producer share and the first-order node degree of the perturbed node (i.e.\  the number of adjacent neighbours) are the network parameters with the highest impact on the survivability. We have found this through  numerical simulations of perturbed multi-node networks in combination with analytic considerations of a two-node approximated network model. Based on these results we can provide the following guidelines for designing a DC power grid capable of surviving the investigated perturbations.
If the voltage survivability limits of a DC power grid are chosen close enough to the equilibrium voltage, in other words, if large voltage deviations should be avoided, a particular producer share maximizes the survivability of the grid. In our particular case ($[\SI{44}{\volt},\SI{48}{\volt}]$) the survivability of simulated network models is optimized for a producer share of $12\,\%$. For producer shares larger than this optimal value, the survivability is enhanced for lower node degrees. This means critical consumers or producers %in a DC power grid with a producer share greater than the optimal value
should be connected to as few neighbouring nodes as possible. If one keeps the real-world constraint of ''$N-1$ stability'' which requires the network to stay connected if a single line breaks down, then the configuration which minimizes the number of neighbouring nodes for all nodes is a ring structure. For a DC power grid with a producer share smaller than the optimal value, the node degree does not affect its survivability, provided that the lower voltage survivability limit lies sufficiently below the band of survived simulations (cf.\ Fig.~\ref{fig:wide}a).
%While this seems to pose a trade-off between implementing redundant features and limiting the number of connections, our analysis shows that redundant first-order bridging edges between neighbouring nodes are not affected by the tested perturbations at all (cf. Fig.~\ref{fig:simulation}). Therefore, redundant bridging edges should be included in the neighbourhood of critical consumer or producer nodes.

We emphasize that our results contrast previous research on the stability of AC power grids. Menck et al.\citep{menck_how_2014} found the average neighbour degree of the perturbed node to have a much stronger influence on the stability of the AC grid than the degree of the perturbed node itself. In contrast, for DC grids, we have not found any distinct influence of the average neighbour degree on the stability. (cf.\ Fig.~\ref{fig:survivability}). However, it is noted that a different measure of stability, survivability instead of the single-node basin stability, has been used in our investigation.

By combining voltage and current limits, the survivability represents an intuitive and overarching but also flexible measure which can be applied to both individual and multitudes of grouped network models. Yet, information regarding the individual voltage and current effects on the network is lost and has to be obtained with different methods. The single-node basin stability is not a suitable stability measure for DC power grids, because there is only one relevant attractor in the investigated parameter regime.

Further research is needed to explore stability measures for power grids in more detail. Since the survivability allows a flexible definition of boundaries, more standardized measures and limit values are required for comparative studies. More heterogeneous network models can be simulated to expand our findings to more realistic scenarios. One possibility is to investigate various control schemes suggested in the literature \citep{bouzid_survey_2015, colak_survey_2015, molzahn_survey_2017} as well as trade-offs between achieving constant power at the consumer nodes and stabilizing behaviour. Moreover, going beyond voltage kicks at single nodes, other perturbation types such as step-like sequences or white voltage (or current) noise remain to be investigated. Our stability analysis is easily extendable to a plethora of additional network measures (including centrality, transitivity and efficiency) or performance measures\cite{paganini}. These suggestions are next steps towards finding the optimal design of stable DC power grids.

\begin{acknowledgments}
J.W., D.E.\ and J.Kr.\ thank the German Academic Scholarship Foundation (Studienstiftung des Deutschen Volkes) for the funding and organisation of the Natur- und Ingenieurwissenschaftliches Kolleg 2017--19 throughout which the research reported here was carried out. We are grateful for stimulating discussions with J. F. Donges, M. Wiedermann, N. Wunderling (PIK) and B. Hecker (LMU Munich) and highly appreciate the kind support from the entire working group.

The work presented was partially funded by the BMBF (Grant No. 03EK3055A) and the Deutsche Forschungsgemeinschaft (Grant No. KU 837/39-1 / RA 516/13-1).
\end{acknowledgments}

%\appendix

%\section{Appendixes}

\section*{References}

%\nocite{*} <<<----- quotes all papers, also the ones which are not cited
\bibliography{manuscript}% Produces the bibliography via BibTeX.

\begin{thebibliography}{45}%
\makeatletter
\providecommand \@ifxundefined [1]{%
 \@ifx{#1\undefined}
}%
\providecommand \@ifnum [1]{%
 \ifnum #1\expandafter \@firstoftwo
 \else \expandafter \@secondoftwo
 \fi
}%
\providecommand \@ifx [1]{%
 \ifx #1\expandafter \@firstoftwo
 \else \expandafter \@secondoftwo
 \fi
}%
\providecommand \natexlab [1]{#1}%
\providecommand \enquote  [1]{``#1''}%
\providecommand \bibnamefont  [1]{#1}%
\providecommand \bibfnamefont [1]{#1}%
\providecommand \citenamefont [1]{#1}%
\providecommand \href@noop [0]{\@secondoftwo}%
\providecommand \href [0]{\begingroup \@sanitize@url \@href}%
\providecommand \@href[1]{\@@startlink{#1}\@@href}%
\providecommand \@@href[1]{\endgroup#1\@@endlink}%
\providecommand \@sanitize@url [0]{\catcode `\\12\catcode `\$12\catcode
  `\&12\catcode `\#12\catcode `\^12\catcode `\_12\catcode `\%12\relax}%
\providecommand \@@startlink[1]{}%
\providecommand \@@endlink[0]{}%
\providecommand \url  [0]{\begingroup\@sanitize@url \@url }%
\providecommand \@url [1]{\endgroup\@href {#1}{\urlprefix }}%
\providecommand \urlprefix  [0]{URL }%
\providecommand \Eprint [0]{\href }%
\providecommand \doibase [0]{http://dx.doi.org/}%
\providecommand \selectlanguage [0]{\@gobble}%
\providecommand \bibinfo  [0]{\@secondoftwo}%
\providecommand \bibfield  [0]{\@secondoftwo}%
\providecommand \translation [1]{[#1]}%
\providecommand \BibitemOpen [0]{}%
\providecommand \bibitemStop [0]{}%
\providecommand \bibitemNoStop [0]{.\EOS\space}%
\providecommand \EOS [0]{\spacefactor3000\relax}%
\providecommand \BibitemShut  [1]{\csname bibitem#1\endcsname}%
\let\auto@bib@innerbib\@empty
%</preamble>
\bibitem [{\citenamefont {McPherson}(2012)}]{mcpherson_war_2012}%
  \BibitemOpen
  \bibfield  {author} {\bibinfo {author} {\bibfnamefont {S.~S.}\ \bibnamefont
  {McPherson}},\ }\href@noop {} {\emph {\bibinfo {title} {War of the
  {Currents}: {Thomas} {Edison} vs {Nikola} {Tesla}}}}\ (\bibinfo  {publisher}
  {Twenty-First Century Books},\ \bibinfo {year} {2012})\BibitemShut {NoStop}%
\bibitem [{\citenamefont {Brinkman}, \citenamefont {Haggan},\ and\
  \citenamefont {Troutman}(1997)}]{brinkman_history_1997}%
  \BibitemOpen
  \bibfield  {author} {\bibinfo {author} {\bibfnamefont {W.~F.}\ \bibnamefont
  {Brinkman}}, \bibinfo {author} {\bibfnamefont {D.~E.}\ \bibnamefont
  {Haggan}}, \ and\ \bibinfo {author} {\bibfnamefont {W.~W.}\ \bibnamefont
  {Troutman}},\ }\href {\doibase 10.1109/4.643644} {\bibfield  {journal}
  {\bibinfo  {journal} {IEEE Journal of Solid-State Circuits}\ }\textbf
  {\bibinfo {volume} {32}},\ \bibinfo {pages} {1858} (\bibinfo {year}
  {1997})}\BibitemShut {NoStop}%
\bibitem [{\citenamefont {Dragicevic}\ \emph {et~al.}(2015)\citenamefont
  {Dragicevic}, \citenamefont {Lu}, \citenamefont {Vasquez},\ and\
  \citenamefont {Guerrero}}]{dragicevic_dc_2015}%
  \BibitemOpen
  \bibfield  {author} {\bibinfo {author} {\bibfnamefont {T.}~\bibnamefont
  {Dragicevic}}, \bibinfo {author} {\bibfnamefont {X.}~\bibnamefont {Lu}},
  \bibinfo {author} {\bibfnamefont {J.}~\bibnamefont {Vasquez}}, \ and\
  \bibinfo {author} {\bibfnamefont {J.}~\bibnamefont {Guerrero}},\ }\href
  {\doibase 10.1109/TPEL.2015.2478859} {\bibfield  {journal} {\bibinfo
  {journal} {IEEE Transactions on Power Electronics}\ ,\ \bibinfo {pages} {1}}
  (\bibinfo {year} {2015})}\BibitemShut {NoStop}%
\bibitem [{\citenamefont {Kakigano}, \citenamefont {Miura},\ and\ \citenamefont
  {Ise}(2010)}]{kakigano_low-voltage_2010}%
  \BibitemOpen
  \bibfield  {author} {\bibinfo {author} {\bibfnamefont {H.}~\bibnamefont
  {Kakigano}}, \bibinfo {author} {\bibfnamefont {Y.}~\bibnamefont {Miura}}, \
  and\ \bibinfo {author} {\bibfnamefont {T.}~\bibnamefont {Ise}},\ }\href
  {\doibase 10.1109/TPEL.2010.2077682} {\bibfield  {journal} {\bibinfo
  {journal} {IEEE Transactions on Power Electronics}\ }\textbf {\bibinfo
  {volume} {25}},\ \bibinfo {pages} {3066} (\bibinfo {year}
  {2010})}\BibitemShut {NoStop}%
\bibitem [{\citenamefont {Grunberg}\ and\ \citenamefont
  {Gayme}(2016)}]{grunberg2016performance}%
  \BibitemOpen
  \bibfield  {author} {\bibinfo {author} {\bibfnamefont {T.~W.}\ \bibnamefont
  {Grunberg}}\ and\ \bibinfo {author} {\bibfnamefont {D.~F.}\ \bibnamefont
  {Gayme}},\ }\href@noop {} {\bibfield  {journal} {\bibinfo  {journal} {IEEE
  Transactions on Control of Network Systems}\ }\textbf {\bibinfo {volume}
  {5}},\ \bibinfo {pages} {456} (\bibinfo {year} {2016})}\BibitemShut {NoStop}%
\bibitem [{\citenamefont {Pagnier}\ and\ \citenamefont
  {Jacquod}(2019)}]{10.1371/journal.pone.0213550}%
  \BibitemOpen
  \bibfield  {author} {\bibinfo {author} {\bibfnamefont {L.}~\bibnamefont
  {Pagnier}}\ and\ \bibinfo {author} {\bibfnamefont {P.}~\bibnamefont
  {Jacquod}},\ }\href {\doibase 10.1371/journal.pone.0213550} {\bibfield
  {journal} {\bibinfo  {journal} {PLOS ONE}\ }\textbf {\bibinfo {volume}
  {14}},\ \bibinfo {pages} {1} (\bibinfo {year} {2019})}\BibitemShut {NoStop}%
\bibitem [{\citenamefont {Poolla}, \citenamefont {Bolognani},\ and\
  \citenamefont {D{\"o}rfler}(2017)}]{poolla2017optimal}%
  \BibitemOpen
  \bibfield  {author} {\bibinfo {author} {\bibfnamefont {B.~K.}\ \bibnamefont
  {Poolla}}, \bibinfo {author} {\bibfnamefont {S.}~\bibnamefont {Bolognani}}, \
  and\ \bibinfo {author} {\bibfnamefont {F.}~\bibnamefont {D{\"o}rfler}},\
  }\href@noop {} {\bibfield  {journal} {\bibinfo  {journal} {IEEE Transactions
  on Automatic Control}\ }\textbf {\bibinfo {volume} {62}},\ \bibinfo {pages}
  {6209} (\bibinfo {year} {2017})}\BibitemShut {NoStop}%
\bibitem [{\citenamefont {Tyloo}, \citenamefont {Coletta},\ and\ \citenamefont
  {Jacquod}(2018)}]{tyloo2018robustness}%
  \BibitemOpen
  \bibfield  {author} {\bibinfo {author} {\bibfnamefont {M.}~\bibnamefont
  {Tyloo}}, \bibinfo {author} {\bibfnamefont {T.}~\bibnamefont {Coletta}}, \
  and\ \bibinfo {author} {\bibfnamefont {P.}~\bibnamefont {Jacquod}},\
  }\href@noop {} {\bibfield  {journal} {\bibinfo  {journal} {Physical review
  letters}\ }\textbf {\bibinfo {volume} {120}},\ \bibinfo {pages} {084101}
  (\bibinfo {year} {2018})}\BibitemShut {NoStop}%
\bibitem [{\citenamefont {Tamrakar}, \citenamefont {Conrath},\ and\
  \citenamefont {Kettemann}(2018)}]{tamrakar2018propagation}%
  \BibitemOpen
  \bibfield  {author} {\bibinfo {author} {\bibfnamefont {S.}~\bibnamefont
  {Tamrakar}}, \bibinfo {author} {\bibfnamefont {M.}~\bibnamefont {Conrath}}, \
  and\ \bibinfo {author} {\bibfnamefont {S.}~\bibnamefont {Kettemann}},\
  }\href@noop {} {\bibfield  {journal} {\bibinfo  {journal} {Scientific
  reports}\ }\textbf {\bibinfo {volume} {8}},\ \bibinfo {pages} {6459}
  (\bibinfo {year} {2018})}\BibitemShut {NoStop}%
\bibitem [{\citenamefont {Guo}, \citenamefont {Zhao},\ and\ \citenamefont
  {Low}(2018)}]{guo2018graph}%
  \BibitemOpen
  \bibfield  {author} {\bibinfo {author} {\bibfnamefont {L.}~\bibnamefont
  {Guo}}, \bibinfo {author} {\bibfnamefont {C.}~\bibnamefont {Zhao}}, \ and\
  \bibinfo {author} {\bibfnamefont {S.~H.}\ \bibnamefont {Low}},\ }in\
  \href@noop {} {\emph {\bibinfo {booktitle} {2018 IEEE Conference on Decision
  and Control (CDC)}}}\ (\bibinfo {organization} {IEEE},\ \bibinfo {year}
  {2018})\ pp.\ \bibinfo {pages} {158--165}\BibitemShut {NoStop}%
\bibitem [{\citenamefont {Zhang}\ \emph {et~al.}(2019)\citenamefont {Zhang},
  \citenamefont {Hallerberg}, \citenamefont {Matthiae}, \citenamefont
  {Witthaut},\ and\ \citenamefont {Timme}}]{Zhangeaav1027}%
  \BibitemOpen
  \bibfield  {author} {\bibinfo {author} {\bibfnamefont {X.}~\bibnamefont
  {Zhang}}, \bibinfo {author} {\bibfnamefont {S.}~\bibnamefont {Hallerberg}},
  \bibinfo {author} {\bibfnamefont {M.}~\bibnamefont {Matthiae}}, \bibinfo
  {author} {\bibfnamefont {D.}~\bibnamefont {Witthaut}}, \ and\ \bibinfo
  {author} {\bibfnamefont {M.}~\bibnamefont {Timme}},\ }\href {\doibase
  10.1126/sciadv.aav1027} {\bibfield  {journal} {\bibinfo  {journal} {Science
  Advances}\ }\textbf {\bibinfo {volume} {5}} (\bibinfo {year} {2019}),\
  10.1126/sciadv.aav1027},\ \Eprint
  {http://arxiv.org/abs/https://advances.sciencemag.org/content/5/7/eaav1027.full.pdf}
  {https://advances.sciencemag.org/content/5/7/eaav1027.full.pdf} \BibitemShut
  {NoStop}%
\bibitem [{\citenamefont {Plietzsch}\ \emph {et~al.}(2019)\citenamefont
  {Plietzsch}, \citenamefont {Auer}, \citenamefont {Kurths},\ and\
  \citenamefont {Hellmann}}]{plietzsch2019generalized}%
  \BibitemOpen
  \bibfield  {author} {\bibinfo {author} {\bibfnamefont {A.}~\bibnamefont
  {Plietzsch}}, \bibinfo {author} {\bibfnamefont {S.}~\bibnamefont {Auer}},
  \bibinfo {author} {\bibfnamefont {J.}~\bibnamefont {Kurths}}, \ and\ \bibinfo
  {author} {\bibfnamefont {F.}~\bibnamefont {Hellmann}},\ }\href@noop {}
  {\bibfield  {journal} {\bibinfo  {journal} {arXiv preprint arXiv:1903.09585}\
  } (\bibinfo {year} {2019})}\BibitemShut {NoStop}%
\bibitem [{\citenamefont {Zhang}, \citenamefont {Ma},\ and\ \citenamefont
  {Timme}(2019)}]{zhang2019dynamic}%
  \BibitemOpen
  \bibfield  {author} {\bibinfo {author} {\bibfnamefont {X.}~\bibnamefont
  {Zhang}}, \bibinfo {author} {\bibfnamefont {C.}~\bibnamefont {Ma}}, \ and\
  \bibinfo {author} {\bibfnamefont {M.}~\bibnamefont {Timme}},\ }\href@noop {}
  {\bibfield  {journal} {\bibinfo  {journal} {arXiv preprint arXiv:1908.00957}\
  } (\bibinfo {year} {2019})}\BibitemShut {NoStop}%
\bibitem [{\citenamefont {Menck}\ \emph {et~al.}(2014)\citenamefont {Menck},
  \citenamefont {Heitzig}, \citenamefont {Kurths},\ and\ \citenamefont
  {Joachim~Schellnhuber}}]{menck_how_2014}%
  \BibitemOpen
  \bibfield  {author} {\bibinfo {author} {\bibfnamefont {P.~J.}\ \bibnamefont
  {Menck}}, \bibinfo {author} {\bibfnamefont {J.}~\bibnamefont {Heitzig}},
  \bibinfo {author} {\bibfnamefont {J.}~\bibnamefont {Kurths}}, \ and\ \bibinfo
  {author} {\bibfnamefont {H.}~\bibnamefont {Joachim~Schellnhuber}},\ }\href
  {\doibase 10.1038/ncomms4969} {\bibfield  {journal} {\bibinfo  {journal}
  {Nature Communications}\ }\textbf {\bibinfo {volume} {5}},\ \bibinfo {pages}
  {3969} (\bibinfo {year} {2014})}\BibitemShut {NoStop}%
\bibitem [{\citenamefont {Schultz}, \citenamefont {Heitzig},\ and\
  \citenamefont {Kurths}(2014{\natexlab{a}})}]{Schultz_2014}%
  \BibitemOpen
  \bibfield  {author} {\bibinfo {author} {\bibfnamefont {P.}~\bibnamefont
  {Schultz}}, \bibinfo {author} {\bibfnamefont {J.}~\bibnamefont {Heitzig}}, \
  and\ \bibinfo {author} {\bibfnamefont {J.}~\bibnamefont {Kurths}},\ }\href
  {\doibase 10.1088/1367-2630/16/12/125001} {\bibfield  {journal} {\bibinfo
  {journal} {New Journal of Physics}\ }\textbf {\bibinfo {volume} {16}},\
  \bibinfo {pages} {125001} (\bibinfo {year} {2014}{\natexlab{a}})}\BibitemShut
  {NoStop}%
\bibitem [{\citenamefont {Auer}\ \emph {et~al.}(2016)\citenamefont {Auer},
  \citenamefont {Kleis}, \citenamefont {Schultz}, \citenamefont {Kurths},\ and\
  \citenamefont {Hellmann}}]{auer2016impact}%
  \BibitemOpen
  \bibfield  {author} {\bibinfo {author} {\bibfnamefont {S.}~\bibnamefont
  {Auer}}, \bibinfo {author} {\bibfnamefont {K.}~\bibnamefont {Kleis}},
  \bibinfo {author} {\bibfnamefont {P.}~\bibnamefont {Schultz}}, \bibinfo
  {author} {\bibfnamefont {J.}~\bibnamefont {Kurths}}, \ and\ \bibinfo {author}
  {\bibfnamefont {F.}~\bibnamefont {Hellmann}},\ }\href@noop {} {\bibfield
  {journal} {\bibinfo  {journal} {The European Physical Journal Special
  Topics}\ }\textbf {\bibinfo {volume} {225}},\ \bibinfo {pages} {609}
  (\bibinfo {year} {2016})}\BibitemShut {NoStop}%
\bibitem [{\citenamefont {Hellmann}\ \emph
  {et~al.}(2016{\natexlab{a}})\citenamefont {Hellmann}, \citenamefont
  {Schultz}, \citenamefont {Grabow}, \citenamefont {Heitzig},\ and\
  \citenamefont {Kurths}}]{hellmann2016survivability}%
  \BibitemOpen
  \bibfield  {author} {\bibinfo {author} {\bibfnamefont {F.}~\bibnamefont
  {Hellmann}}, \bibinfo {author} {\bibfnamefont {P.}~\bibnamefont {Schultz}},
  \bibinfo {author} {\bibfnamefont {C.}~\bibnamefont {Grabow}}, \bibinfo
  {author} {\bibfnamefont {J.}~\bibnamefont {Heitzig}}, \ and\ \bibinfo
  {author} {\bibfnamefont {J.}~\bibnamefont {Kurths}},\ }\href@noop {}
  {\bibfield  {journal} {\bibinfo  {journal} {Scientific reports}\ }\textbf
  {\bibinfo {volume} {6}},\ \bibinfo {pages} {29654} (\bibinfo {year}
  {2016}{\natexlab{a}})}\BibitemShut {NoStop}%
\bibitem [{\citenamefont {Nitzbon}\ \emph
  {et~al.}(2017{\natexlab{a}})\citenamefont {Nitzbon}, \citenamefont {Schultz},
  \citenamefont {Heitzig}, \citenamefont {Kurths},\ and\ \citenamefont
  {Hellmann}}]{nitzbon2017deciphering}%
  \BibitemOpen
  \bibfield  {author} {\bibinfo {author} {\bibfnamefont {J.}~\bibnamefont
  {Nitzbon}}, \bibinfo {author} {\bibfnamefont {P.}~\bibnamefont {Schultz}},
  \bibinfo {author} {\bibfnamefont {J.}~\bibnamefont {Heitzig}}, \bibinfo
  {author} {\bibfnamefont {J.}~\bibnamefont {Kurths}}, \ and\ \bibinfo {author}
  {\bibfnamefont {F.}~\bibnamefont {Hellmann}},\ }\href@noop {} {\bibfield
  {journal} {\bibinfo  {journal} {New Journal of Physics}\ }\textbf {\bibinfo
  {volume} {19}},\ \bibinfo {pages} {033029} (\bibinfo {year}
  {2017}{\natexlab{a}})}\BibitemShut {NoStop}%
\bibitem [{\citenamefont {Auer}\ \emph {et~al.}(2017)\citenamefont {Auer},
  \citenamefont {Hellmann}, \citenamefont {Krause},\ and\ \citenamefont
  {Kurths}}]{auer2017stability}%
  \BibitemOpen
  \bibfield  {author} {\bibinfo {author} {\bibfnamefont {S.}~\bibnamefont
  {Auer}}, \bibinfo {author} {\bibfnamefont {F.}~\bibnamefont {Hellmann}},
  \bibinfo {author} {\bibfnamefont {M.}~\bibnamefont {Krause}}, \ and\ \bibinfo
  {author} {\bibfnamefont {J.}~\bibnamefont {Kurths}},\ }\href@noop {}
  {\bibfield  {journal} {\bibinfo  {journal} {Chaos: An Interdisciplinary
  Journal of Nonlinear Science}\ }\textbf {\bibinfo {volume} {27}},\ \bibinfo
  {pages} {127003} (\bibinfo {year} {2017})}\BibitemShut {NoStop}%
\bibitem [{\citenamefont {Hellmann}\ \emph {et~al.}(2018)\citenamefont
  {Hellmann}, \citenamefont {Schultz}, \citenamefont {Jaros}, \citenamefont
  {Levchenko}, \citenamefont {Kapitaniak}, \citenamefont {Kurths},\ and\
  \citenamefont {Maistrenko}}]{hellmann2018network}%
  \BibitemOpen
  \bibfield  {author} {\bibinfo {author} {\bibfnamefont {F.}~\bibnamefont
  {Hellmann}}, \bibinfo {author} {\bibfnamefont {P.}~\bibnamefont {Schultz}},
  \bibinfo {author} {\bibfnamefont {P.}~\bibnamefont {Jaros}}, \bibinfo
  {author} {\bibfnamefont {R.}~\bibnamefont {Levchenko}}, \bibinfo {author}
  {\bibfnamefont {T.}~\bibnamefont {Kapitaniak}}, \bibinfo {author}
  {\bibfnamefont {J.}~\bibnamefont {Kurths}}, \ and\ \bibinfo {author}
  {\bibfnamefont {Y.}~\bibnamefont {Maistrenko}},\ }\href@noop {} {\bibfield
  {journal} {\bibinfo  {journal} {arXiv preprint arXiv:1811.11518}\ } (\bibinfo
  {year} {2018})}\BibitemShut {NoStop}%
\bibitem [{\citenamefont {Kim}, \citenamefont {Lee},\ and\ \citenamefont
  {Holme}(2016)}]{kim2016building}%
  \BibitemOpen
  \bibfield  {author} {\bibinfo {author} {\bibfnamefont {H.}~\bibnamefont
  {Kim}}, \bibinfo {author} {\bibfnamefont {S.~H.}\ \bibnamefont {Lee}}, \ and\
  \bibinfo {author} {\bibfnamefont {P.}~\bibnamefont {Holme}},\ }\href@noop {}
  {\bibfield  {journal} {\bibinfo  {journal} {Physical Review E}\ }\textbf
  {\bibinfo {volume} {93}},\ \bibinfo {pages} {062318} (\bibinfo {year}
  {2016})}\BibitemShut {NoStop}%
\bibitem [{\citenamefont {Wolff}, \citenamefont {Lind},\ and\ \citenamefont
  {Maass}(2018)}]{wolff2018power}%
  \BibitemOpen
  \bibfield  {author} {\bibinfo {author} {\bibfnamefont {M.~F.}\ \bibnamefont
  {Wolff}}, \bibinfo {author} {\bibfnamefont {P.~G.}\ \bibnamefont {Lind}}, \
  and\ \bibinfo {author} {\bibfnamefont {P.}~\bibnamefont {Maass}},\
  }\href@noop {} {\bibfield  {journal} {\bibinfo  {journal} {Chaos: An
  Interdisciplinary Journal of Nonlinear Science}\ }\textbf {\bibinfo {volume}
  {28}},\ \bibinfo {pages} {103120} (\bibinfo {year} {2018})}\BibitemShut
  {NoStop}%
\bibitem [{\citenamefont {Schultz}, \citenamefont {Heitzig},\ and\
  \citenamefont {Kurths}(2014{\natexlab{b}})}]{Schultz2014growth}%
  \BibitemOpen
  \bibfield  {author} {\bibinfo {author} {\bibfnamefont {P.}~\bibnamefont
  {Schultz}}, \bibinfo {author} {\bibfnamefont {J.}~\bibnamefont {Heitzig}}, \
  and\ \bibinfo {author} {\bibfnamefont {J.}~\bibnamefont {Kurths}},\ }\href
  {\doibase 10.1140/epjst/e2014-02279-6} {\bibfield  {journal} {\bibinfo
  {journal} {The European Physical Journal Special Topics}\ }\textbf {\bibinfo
  {volume} {223}},\ \bibinfo {pages} {2593} (\bibinfo {year}
  {2014}{\natexlab{b}})}\BibitemShut {NoStop}%
\bibitem [{\citenamefont {Dragicevic}\ \emph {et~al.}(2016)\citenamefont
  {Dragicevic}, \citenamefont {Lu}, \citenamefont {Vasquez},\ and\
  \citenamefont {Guerrero}}]{dragicevic_dc_2016}%
  \BibitemOpen
  \bibfield  {author} {\bibinfo {author} {\bibfnamefont {T.}~\bibnamefont
  {Dragicevic}}, \bibinfo {author} {\bibfnamefont {X.}~\bibnamefont {Lu}},
  \bibinfo {author} {\bibfnamefont {J.~C.}\ \bibnamefont {Vasquez}}, \ and\
  \bibinfo {author} {\bibfnamefont {J.~M.}\ \bibnamefont {Guerrero}},\ }\href
  {\doibase 10.1109/TPEL.2015.2464277} {\bibfield  {journal} {\bibinfo
  {journal} {IEEE Transactions on Power Electronics}\ }\textbf {\bibinfo
  {volume} {31}},\ \bibinfo {pages} {3528} (\bibinfo {year}
  {2016})}\BibitemShut {NoStop}%
\bibitem [{\citenamefont {Zhao}\ and\ \citenamefont
  {D{\"o}rfler}(2015)}]{zhao2015distributed}%
  \BibitemOpen
  \bibfield  {author} {\bibinfo {author} {\bibfnamefont {J.}~\bibnamefont
  {Zhao}}\ and\ \bibinfo {author} {\bibfnamefont {F.}~\bibnamefont
  {D{\"o}rfler}},\ }\href@noop {} {\bibfield  {journal} {\bibinfo  {journal}
  {Automatica}\ }\textbf {\bibinfo {volume} {61}},\ \bibinfo {pages} {18}
  (\bibinfo {year} {2015})}\BibitemShut {NoStop}%
\bibitem [{\citenamefont {Meng}\ \emph {et~al.}(2017)\citenamefont {Meng},
  \citenamefont {Shafiee}, \citenamefont {Trecate}, \citenamefont {Karimi},
  \citenamefont {Fulwani}, \citenamefont {Lu},\ and\ \citenamefont
  {Guerrero}}]{meng2017review}%
  \BibitemOpen
  \bibfield  {author} {\bibinfo {author} {\bibfnamefont {L.}~\bibnamefont
  {Meng}}, \bibinfo {author} {\bibfnamefont {Q.}~\bibnamefont {Shafiee}},
  \bibinfo {author} {\bibfnamefont {G.~F.}\ \bibnamefont {Trecate}}, \bibinfo
  {author} {\bibfnamefont {H.}~\bibnamefont {Karimi}}, \bibinfo {author}
  {\bibfnamefont {D.}~\bibnamefont {Fulwani}}, \bibinfo {author} {\bibfnamefont
  {X.}~\bibnamefont {Lu}}, \ and\ \bibinfo {author} {\bibfnamefont {J.~M.}\
  \bibnamefont {Guerrero}},\ }\href@noop {} {\bibfield  {journal} {\bibinfo
  {journal} {IEEE Journal of Emerging and Selected Topics in Power
  Electronics}\ }\textbf {\bibinfo {volume} {5}},\ \bibinfo {pages} {928}
  (\bibinfo {year} {2017})}\BibitemShut {NoStop}%
\bibitem [{\citenamefont {De~Persis}, \citenamefont {Weitenberg},\ and\
  \citenamefont {D{\"o}rfler}(2018)}]{de2018power}%
  \BibitemOpen
  \bibfield  {author} {\bibinfo {author} {\bibfnamefont {C.}~\bibnamefont
  {De~Persis}}, \bibinfo {author} {\bibfnamefont {E.~R.}\ \bibnamefont
  {Weitenberg}}, \ and\ \bibinfo {author} {\bibfnamefont {F.}~\bibnamefont
  {D{\"o}rfler}},\ }\href@noop {} {\bibfield  {journal} {\bibinfo  {journal}
  {Automatica}\ }\textbf {\bibinfo {volume} {89}},\ \bibinfo {pages} {364}
  (\bibinfo {year} {2018})}\BibitemShut {NoStop}%
\bibitem [{\citenamefont {Cucuzzella}\ \emph {et~al.}(2018)\citenamefont
  {Cucuzzella}, \citenamefont {Trip}, \citenamefont {De~Persis}, \citenamefont
  {Cheng}, \citenamefont {Ferrara},\ and\ \citenamefont {van~der
  Schaft}}]{cucuzzella2018robust}%
  \BibitemOpen
  \bibfield  {author} {\bibinfo {author} {\bibfnamefont {M.}~\bibnamefont
  {Cucuzzella}}, \bibinfo {author} {\bibfnamefont {S.}~\bibnamefont {Trip}},
  \bibinfo {author} {\bibfnamefont {C.}~\bibnamefont {De~Persis}}, \bibinfo
  {author} {\bibfnamefont {X.}~\bibnamefont {Cheng}}, \bibinfo {author}
  {\bibfnamefont {A.}~\bibnamefont {Ferrara}}, \ and\ \bibinfo {author}
  {\bibfnamefont {A.}~\bibnamefont {van~der Schaft}},\ }\href@noop {}
  {\bibfield  {journal} {\bibinfo  {journal} {IEEE Transactions on Control
  Systems Technology}\ }\textbf {\bibinfo {volume} {27}},\ \bibinfo {pages}
  {1583} (\bibinfo {year} {2018})}\BibitemShut {NoStop}%
\bibitem [{\citenamefont {Trip}\ \emph {et~al.}(2019)\citenamefont {Trip},
  \citenamefont {Cucuzzella}, \citenamefont {Cheng},\ and\ \citenamefont
  {Scherpen}}]{trip2019distributed}%
  \BibitemOpen
  \bibfield  {author} {\bibinfo {author} {\bibfnamefont {S.}~\bibnamefont
  {Trip}}, \bibinfo {author} {\bibfnamefont {M.}~\bibnamefont {Cucuzzella}},
  \bibinfo {author} {\bibfnamefont {X.}~\bibnamefont {Cheng}}, \ and\ \bibinfo
  {author} {\bibfnamefont {J.}~\bibnamefont {Scherpen}},\ }\href@noop {}
  {\bibfield  {journal} {\bibinfo  {journal} {IEEE Control Systems Letters}\
  }\textbf {\bibinfo {volume} {3}},\ \bibinfo {pages} {174} (\bibinfo {year}
  {2019})}\BibitemShut {NoStop}%
\bibitem [{\citenamefont {Andreasson}\ \emph {et~al.}(2014)\citenamefont
  {Andreasson}, \citenamefont {Nazari}, \citenamefont {Dimarogonas},
  \citenamefont {Sandberg}, \citenamefont {Johansson},\ and\ \citenamefont
  {Ghandhari}}]{andreasson2014distributed}%
  \BibitemOpen
  \bibfield  {author} {\bibinfo {author} {\bibfnamefont {M.}~\bibnamefont
  {Andreasson}}, \bibinfo {author} {\bibfnamefont {M.}~\bibnamefont {Nazari}},
  \bibinfo {author} {\bibfnamefont {D.~V.}\ \bibnamefont {Dimarogonas}},
  \bibinfo {author} {\bibfnamefont {H.}~\bibnamefont {Sandberg}}, \bibinfo
  {author} {\bibfnamefont {K.~H.}\ \bibnamefont {Johansson}}, \ and\ \bibinfo
  {author} {\bibfnamefont {M.}~\bibnamefont {Ghandhari}},\ }\href@noop {}
  {\bibfield  {journal} {\bibinfo  {journal} {IFAC Proceedings Volumes}\
  }\textbf {\bibinfo {volume} {47}},\ \bibinfo {pages} {11910} (\bibinfo {year}
  {2014})}\BibitemShut {NoStop}%
\bibitem [{\citenamefont {Zonetti}, \citenamefont {Ortega},\ and\ \citenamefont
  {Schiffer}(2017)}]{zonetti2017tool}%
  \BibitemOpen
  \bibfield  {author} {\bibinfo {author} {\bibfnamefont {D.}~\bibnamefont
  {Zonetti}}, \bibinfo {author} {\bibfnamefont {R.}~\bibnamefont {Ortega}}, \
  and\ \bibinfo {author} {\bibfnamefont {J.}~\bibnamefont {Schiffer}},\
  }\href@noop {} {\bibfield  {journal} {\bibinfo  {journal} {IEEE Transactions
  on Control of Network Systems}\ }\textbf {\bibinfo {volume} {5}},\ \bibinfo
  {pages} {1110} (\bibinfo {year} {2017})}\BibitemShut {NoStop}%
\bibitem [{\citenamefont {Anand}\ and\ \citenamefont
  {Fernandes}(2012)}]{anand2012reduced}%
  \BibitemOpen
  \bibfield  {author} {\bibinfo {author} {\bibfnamefont {S.}~\bibnamefont
  {Anand}}\ and\ \bibinfo {author} {\bibfnamefont {B.}~\bibnamefont
  {Fernandes}},\ }\href@noop {} {\bibfield  {journal} {\bibinfo  {journal}
  {IEEE Transactions on Industrial Electronics}\ }\textbf {\bibinfo {volume}
  {60}},\ \bibinfo {pages} {5040} (\bibinfo {year} {2012})}\BibitemShut
  {NoStop}%
\bibitem [{\citenamefont {Groh}\ \emph {et~al.}(2015)\citenamefont {Groh},
  \citenamefont {Philipp}, \citenamefont {Lasch},\ and\ \citenamefont
  {Kirchhoff}}]{swarm2015}%
  \BibitemOpen
  \bibfield  {author} {\bibinfo {author} {\bibfnamefont {S.}~\bibnamefont
  {Groh}}, \bibinfo {author} {\bibfnamefont {D.}~\bibnamefont {Philipp}},
  \bibinfo {author} {\bibfnamefont {B.~E.}\ \bibnamefont {Lasch}}, \ and\
  \bibinfo {author} {\bibfnamefont {H.}~\bibnamefont {Kirchhoff}},\ }in\
  \href@noop {} {\emph {\bibinfo {booktitle} {Decentralized Solutions for
  Developing Economies}}},\ \bibinfo {editor} {edited by\ \bibinfo {editor}
  {\bibfnamefont {S.}~\bibnamefont {Groh}}, \bibinfo {editor} {\bibfnamefont
  {J.}~\bibnamefont {van~der Straeten}}, \bibinfo {editor} {\bibfnamefont
  {B.}~\bibnamefont {Edlefsen~Lasch}}, \bibinfo {editor} {\bibfnamefont
  {D.}~\bibnamefont {Gershenson}}, \bibinfo {editor} {\bibfnamefont
  {W.}~\bibnamefont {Leal~Filho}}, \ and\ \bibinfo {editor} {\bibfnamefont
  {D.~M.}\ \bibnamefont {Kammen}}}\ (\bibinfo  {publisher} {Springer
  International Publishing},\ \bibinfo {address} {Cham},\ \bibinfo {year}
  {2015})\ pp.\ \bibinfo {pages} {3--22}\BibitemShut {NoStop}%
\bibitem [{\citenamefont {Strenge}\ \emph {et~al.}(2017)\citenamefont
  {Strenge}, \citenamefont {Kirchhoff}, \citenamefont {Ndow},\ and\
  \citenamefont {Hellmann}}]{strenge_stability_2017}%
  \BibitemOpen
  \bibfield  {author} {\bibinfo {author} {\bibfnamefont {L.}~\bibnamefont
  {Strenge}}, \bibinfo {author} {\bibfnamefont {H.}~\bibnamefont {Kirchhoff}},
  \bibinfo {author} {\bibfnamefont {G.~L.}\ \bibnamefont {Ndow}}, \ and\
  \bibinfo {author} {\bibfnamefont {F.}~\bibnamefont {Hellmann}},\ }in\ \href
  {\doibase 10.1109/ICDCM.2017.8001040} {\emph {\bibinfo {booktitle} {2017
  {IEEE} {Second} {International} {Conference} on {DC} {Microgrids}
  ({ICDCM})}}}\ (\bibinfo {year} {2017})\ pp.\ \bibinfo {pages}
  {175--180}\BibitemShut {NoStop}%
\bibitem [{\citenamefont {Watts}\ and\ \citenamefont
  {Strogatz}(1998)}]{watts_collective_1998}%
  \BibitemOpen
  \bibfield  {author} {\bibinfo {author} {\bibfnamefont {D.~J.}\ \bibnamefont
  {Watts}}\ and\ \bibinfo {author} {\bibfnamefont {S.~H.}\ \bibnamefont
  {Strogatz}},\ }\href {\doibase 10.1038/30918} {\bibfield  {journal} {\bibinfo
   {journal} {Nature}\ }\textbf {\bibinfo {volume} {393}},\ \bibinfo {pages}
  {440} (\bibinfo {year} {1998})}\BibitemShut {NoStop}%
\bibitem [{\citenamefont {Hellmann}\ \emph
  {et~al.}(2016{\natexlab{b}})\citenamefont {Hellmann}, \citenamefont
  {Schultz}, \citenamefont {Grabow}, \citenamefont {Heitzig},\ and\
  \citenamefont {Kurths}}]{hellmann_survivability_2016}%
  \BibitemOpen
  \bibfield  {author} {\bibinfo {author} {\bibfnamefont {F.}~\bibnamefont
  {Hellmann}}, \bibinfo {author} {\bibfnamefont {P.}~\bibnamefont {Schultz}},
  \bibinfo {author} {\bibfnamefont {C.}~\bibnamefont {Grabow}}, \bibinfo
  {author} {\bibfnamefont {J.}~\bibnamefont {Heitzig}}, \ and\ \bibinfo
  {author} {\bibfnamefont {J.}~\bibnamefont {Kurths}},\ }\href {\doibase
  10.1038/srep29654} {\bibfield  {journal} {\bibinfo  {journal} {Scientific
  Reports}\ }\textbf {\bibinfo {volume} {6}},\ \bibinfo {pages} {29654}
  (\bibinfo {year} {2016}{\natexlab{b}})}\BibitemShut {NoStop}%
\bibitem [{\citenamefont {Singh}, \citenamefont {Gautam},\ and\ \citenamefont
  {Fulwani}(2017)}]{singh_constant_2017}%
  \BibitemOpen
  \bibfield  {author} {\bibinfo {author} {\bibfnamefont {S.}~\bibnamefont
  {Singh}}, \bibinfo {author} {\bibfnamefont {A.~R.}\ \bibnamefont {Gautam}}, \
  and\ \bibinfo {author} {\bibfnamefont {D.}~\bibnamefont {Fulwani}},\ }\href
  {\doibase 10.1016/j.rser.2017.01.027} {\bibfield  {journal} {\bibinfo
  {journal} {Renewable and Sustainable Energy Reviews}\ }\textbf {\bibinfo
  {volume} {72}},\ \bibinfo {pages} {407} (\bibinfo {year} {2017})}\BibitemShut
  {NoStop}%
\bibitem [{\citenamefont {Gavriluta}\ \emph {et~al.}(2014)\citenamefont
  {Gavriluta}, \citenamefont {Candela}, \citenamefont {Citro}, \citenamefont
  {Rocabert}, \citenamefont {Luna},\ and\ \citenamefont
  {Rodr{\'\i}guez}}]{gavriluta2014decentralized}%
  \BibitemOpen
  \bibfield  {author} {\bibinfo {author} {\bibfnamefont {C.}~\bibnamefont
  {Gavriluta}}, \bibinfo {author} {\bibfnamefont {J.~I.}\ \bibnamefont
  {Candela}}, \bibinfo {author} {\bibfnamefont {C.}~\bibnamefont {Citro}},
  \bibinfo {author} {\bibfnamefont {J.}~\bibnamefont {Rocabert}}, \bibinfo
  {author} {\bibfnamefont {A.}~\bibnamefont {Luna}}, \ and\ \bibinfo {author}
  {\bibfnamefont {P.}~\bibnamefont {Rodr{\'\i}guez}},\ }\href@noop {}
  {\bibfield  {journal} {\bibinfo  {journal} {IEEE Transactions on Industry
  Applications}\ }\textbf {\bibinfo {volume} {50}},\ \bibinfo {pages} {4122}
  (\bibinfo {year} {2014})}\BibitemShut {NoStop}%
\bibitem [{\citenamefont {Kozhaya}, \citenamefont {Nassif},\ and\ \citenamefont
  {Najm}(2002)}]{kozhaya_multigrid-like_2002}%
  \BibitemOpen
  \bibfield  {author} {\bibinfo {author} {\bibfnamefont {J.~N.}\ \bibnamefont
  {Kozhaya}}, \bibinfo {author} {\bibfnamefont {S.~R.}\ \bibnamefont {Nassif}},
  \ and\ \bibinfo {author} {\bibfnamefont {F.~N.}\ \bibnamefont {Najm}},\
  }\href {\doibase 10.1109/TCAD.2002.802271} {\bibfield  {journal} {\bibinfo
  {journal} {IEEE Transactions on Computer-Aided Design of Integrated Circuits
  and Systems}\ }\textbf {\bibinfo {volume} {21}},\ \bibinfo {pages} {1148}
  (\bibinfo {year} {2002})}\BibitemShut {NoStop}%
\bibitem [{\citenamefont {Backhaus}\ \emph {et~al.}(2015)\citenamefont
  {Backhaus}, \citenamefont {Swift}, \citenamefont {Chatzivasileiadis},
  \citenamefont {Tschudi}, \citenamefont {Glover}, \citenamefont {Starke},
  \citenamefont {Wang}, \citenamefont {Yue},\ and\ \citenamefont
  {Hammerstrom}}]{backhaus_dc_2015}%
  \BibitemOpen
  \bibfield  {author} {\bibinfo {author} {\bibfnamefont {S.~N.}\ \bibnamefont
  {Backhaus}}, \bibinfo {author} {\bibfnamefont {G.~W.}\ \bibnamefont {Swift}},
  \bibinfo {author} {\bibfnamefont {S.}~\bibnamefont {Chatzivasileiadis}},
  \bibinfo {author} {\bibfnamefont {W.}~\bibnamefont {Tschudi}}, \bibinfo
  {author} {\bibfnamefont {S.}~\bibnamefont {Glover}}, \bibinfo {author}
  {\bibfnamefont {M.}~\bibnamefont {Starke}}, \bibinfo {author} {\bibfnamefont
  {J.}~\bibnamefont {Wang}}, \bibinfo {author} {\bibfnamefont {M.}~\bibnamefont
  {Yue}}, \ and\ \bibinfo {author} {\bibfnamefont {D.}~\bibnamefont
  {Hammerstrom}},\ }\href {\doibase 10.2172/1209276} {\enquote {\bibinfo
  {title} {{DC} {Microgrids} {Scoping} {Study}. {Estimate} of {Technical} and
  {Economic} {Benefits}},}\ }\bibinfo {type} {Tech. Rep.}\ \bibinfo {number}
  {LA-UR--15-22097, 1209276}\ (\bibinfo {year} {2015})\BibitemShut {NoStop}%
\bibitem [{\citenamefont {Nitzbon}\ \emph
  {et~al.}(2017{\natexlab{b}})\citenamefont {Nitzbon}, \citenamefont {Schultz},
  \citenamefont {Heitzig}, \citenamefont {Kurths},\ and\ \citenamefont
  {Hellmann}}]{nitzbon_deciphering_2017}%
  \BibitemOpen
  \bibfield  {author} {\bibinfo {author} {\bibfnamefont {J.}~\bibnamefont
  {Nitzbon}}, \bibinfo {author} {\bibfnamefont {P.}~\bibnamefont {Schultz}},
  \bibinfo {author} {\bibfnamefont {J.}~\bibnamefont {Heitzig}}, \bibinfo
  {author} {\bibfnamefont {J.}~\bibnamefont {Kurths}}, \ and\ \bibinfo {author}
  {\bibfnamefont {F.}~\bibnamefont {Hellmann}},\ }\href {\doibase
  10.1088/1367-2630/aa6321} {\bibfield  {journal} {\bibinfo  {journal} {New
  Journal of Physics}\ }\textbf {\bibinfo {volume} {19}},\ \bibinfo {pages}
  {033029} (\bibinfo {year} {2017}{\natexlab{b}})}\BibitemShut {NoStop}%
\bibitem [{\citenamefont {Bouzid}\ \emph {et~al.}(2015)\citenamefont {Bouzid},
  \citenamefont {Guerrero}, \citenamefont {Cheriti}, \citenamefont {Bouhamida},
  \citenamefont {Sicard},\ and\ \citenamefont
  {Benghanem}}]{bouzid_survey_2015}%
  \BibitemOpen
  \bibfield  {author} {\bibinfo {author} {\bibfnamefont {A.~M.}\ \bibnamefont
  {Bouzid}}, \bibinfo {author} {\bibfnamefont {J.~M.}\ \bibnamefont
  {Guerrero}}, \bibinfo {author} {\bibfnamefont {A.}~\bibnamefont {Cheriti}},
  \bibinfo {author} {\bibfnamefont {M.}~\bibnamefont {Bouhamida}}, \bibinfo
  {author} {\bibfnamefont {P.}~\bibnamefont {Sicard}}, \ and\ \bibinfo {author}
  {\bibfnamefont {M.}~\bibnamefont {Benghanem}},\ }\href {\doibase
  10.1016/j.rser.2015.01.016} {\bibfield  {journal} {\bibinfo  {journal}
  {Renewable and Sustainable Energy Reviews}\ }\textbf {\bibinfo {volume}
  {44}},\ \bibinfo {pages} {751} (\bibinfo {year} {2015})}\BibitemShut
  {NoStop}%
\bibitem [{\citenamefont {Colak}\ \emph {et~al.}(2015)\citenamefont {Colak},
  \citenamefont {Kabalci}, \citenamefont {Fulli},\ and\ \citenamefont
  {Lazarou}}]{colak_survey_2015}%
  \BibitemOpen
  \bibfield  {author} {\bibinfo {author} {\bibfnamefont {I.}~\bibnamefont
  {Colak}}, \bibinfo {author} {\bibfnamefont {E.}~\bibnamefont {Kabalci}},
  \bibinfo {author} {\bibfnamefont {G.}~\bibnamefont {Fulli}}, \ and\ \bibinfo
  {author} {\bibfnamefont {S.}~\bibnamefont {Lazarou}},\ }\href {\doibase
  10.1016/j.rser.2015.03.031} {\bibfield  {journal} {\bibinfo  {journal}
  {Renewable and Sustainable Energy Reviews}\ }\textbf {\bibinfo {volume}
  {47}},\ \bibinfo {pages} {562} (\bibinfo {year} {2015})}\BibitemShut
  {NoStop}%
\bibitem [{\citenamefont {Molzahn}\ \emph {et~al.}(2017)\citenamefont
  {Molzahn}, \citenamefont {Dörfler}, \citenamefont {Sandberg}, \citenamefont
  {Low}, \citenamefont {Chakrabarti}, \citenamefont {Baldick},\ and\
  \citenamefont {Lavaei}}]{molzahn_survey_2017}%
  \BibitemOpen
  \bibfield  {author} {\bibinfo {author} {\bibfnamefont {D.~K.}\ \bibnamefont
  {Molzahn}}, \bibinfo {author} {\bibfnamefont {F.}~\bibnamefont {Dörfler}},
  \bibinfo {author} {\bibfnamefont {H.}~\bibnamefont {Sandberg}}, \bibinfo
  {author} {\bibfnamefont {S.~H.}\ \bibnamefont {Low}}, \bibinfo {author}
  {\bibfnamefont {S.}~\bibnamefont {Chakrabarti}}, \bibinfo {author}
  {\bibfnamefont {R.}~\bibnamefont {Baldick}}, \ and\ \bibinfo {author}
  {\bibfnamefont {J.}~\bibnamefont {Lavaei}},\ }\href {\doibase
  10.1109/TSG.2017.2720471} {\bibfield  {journal} {\bibinfo  {journal} {IEEE
  Transactions on Smart Grid}\ }\textbf {\bibinfo {volume} {8}},\ \bibinfo
  {pages} {2941} (\bibinfo {year} {2017})}\BibitemShut {NoStop}%
\bibitem [{\citenamefont {{Paganini}}\ and\ \citenamefont
  {{Mallada}}(2017)}]{paganini}%
  \BibitemOpen
  \bibfield  {author} {\bibinfo {author} {\bibfnamefont {F.}~\bibnamefont
  {{Paganini}}}\ and\ \bibinfo {author} {\bibfnamefont {E.}~\bibnamefont
  {{Mallada}}},\ }\bibfield  {booktitle} {\emph {\bibinfo {booktitle} {2017
  55th Annual Allerton Conference on Communication, Control, and Computing
  (Allerton)}},\ }\href {\doibase 10.1109/ALLERTON.2017.8262755} {\ ,\ \bibinfo
  {pages} {324} (\bibinfo {year} {2017})}\BibitemShut {NoStop}%
\end{thebibliography}%

\end{document}